\documentclass[aps,prb,reprint,amsmath,amssymb,superscriptaddress]{revtex4-2}
\usepackage{graphicx}
\usepackage{epstopdf}
\usepackage{amsmath}
\usepackage{dutchcal}
\usepackage{hyperref}
\usepackage{appendix}
\usepackage[capitalize]{cleveref}
\usepackage{subfigure}
\usepackage{indentfirst}
\usepackage{parskip}
\usepackage{verbatim}
\hypersetup{hypertex=true,
			colorlinks=true,
			linkcolor=blue,
			anchorcolor=blue,
			urlcolor=blue,
			citecolor=blue}
\usepackage{amsmath,lipsum}
\newcommand{\myref}[1]{Eq.\ref{#1}}
\begin{document}

\title{Auxiliary dynamical mean-field approach for Anderson-Hubbard model with off-diagonal disorder}
\author{Zelei Zhang}
\affiliation{School of physical science and technology, ShanghaiTech University, Shanghai, 201210, China}
\author{Jiawei Yan}
\affiliation{2020 X-LAB, Shanghai Institute of Microsystem and Information Technology, Chinese Academy of Sciences, Shanghai 200050, China}
\author{Li Huang}
\affiliation{Institute of Materials, China Academy of Engineering Physics, Sichuan Jiangyou 621900, China}
\author{Youqi Ke }
\email{keyq@shanghaitech.edu.cn}
\affiliation{School of physical science and technology, ShanghaiTech University, Shanghai, 201210, China}

\date{\today}

\begin{abstract}
This work reports a theoretical framework that combines the auxiliary coherent potential approximation (ACPA-DMFT) with dynamical mean-field theory to study strongly correlated and disordered electronic systems with both diagonal and off-diagonal disorders. In this method, by introducing an auxiliary coupling space with extended local degree of freedom,the diagonal and off-diagonal disorders are treated in a unified and self-consistent framework of coherent potential approximation, within which the dynamical mean-field theory is naturally combined to handle the strongly correlated Anderson-Hubbard model. By using this approach, we compute matsubara Green’s functions for a simple cubic lattice at finite temperatures and derive impurity spectral functions through the maximum entropy method. Our results reveal the critical influence of off-diagonal disorder on Mott-type metal-insulator transitions. Specifically, a reentrant phenomenon is identified, where the system transitions between insulating and metallic states under varying interaction strengths. The ACPA-DMFT method provides an efficient and robust computational method for exploring the intricate interplay of disorder and strong correlations.  
\end{abstract}

\maketitle

\section{Introduction}

Disorders and electron correlations are two critical factors in tuning the electronic structure of modern functional materials. 
Disorders in crystalline materials, manifested as multiple electron scattering off the impurities, can lead to Anderson localization \cite{Disorder1985, AndersonTransitions2008}.
Meanwhile, electron correlations, characterized by electron-electron interactions, can  destabilize the Fermi surface and induce Mott-type metal-insulator transitions (MIT) \cite{MITRMP1998, MITmott1990, Mott_1949}. 
The interplay between these two factors can give rise to emergent phenomena, such as the competition between Mott and Anderson MIT \cite{Andersonloc50},  disorder-driven transitions \cite{ZhuPRL2019}, disorder supressed or enhanced unconventional superconductivities \cite{disorderSC,SPSC,SPSC2022}, and disorder-inhibit in the fractional quantum Hall effect \cite{FQHE1984}. Studies of the Anderson-Hubbard model, especially in the presence of diagonal disorder, have revealed reentrance behavior under certain conditions \cite{Lombardo2006}, challenging the conventional view of unidirectional phase transitions between metal and insulator states.  

Theoretical treatment of either disorder or strong electron correlations poses significant challenges. 
These difficulties arise from the exponential growth of solution spaces associated with each problem.
For disordered materials, the difficulty stems from the vast number of atomic configurations, each breaking spatial invariant symmetry that necessitating the use of large supercells to approximate symmetry restoration.
While in the correlated systems, the challenge lies in the expotentially grows of the Hilbert space due to the entanglement of interacting electrons that one has to integrated out those irrelavanet degrees of freedom.
Consequently, exact diagonalization methods are limited to small clusters in the disordered interacting electronic system, thus necessitating the development of advanced computational approaches to address realistic material systems.

Dynamical mean field theory (DMFT) \cite{DMFT1989, DMFTJarrell1992, Georges1992, DMFTRMP1996, DMFTRMP2006} is widely used to predict the electronic structure of strongly correlated materials. 
Its central idea lies in mapping the quantum lattice model to an impurity model, which then can be solved numerically \cite{RMPGeorges1996, RMPKotliar2006, Hartmann1989}.
For the disordered cases, the coherent potential approximation (CPA) \cite{gonis1992green, CPA1974, CPASoven1967, CPATaylor1967, CPA1968} shares the same idea to DMFT.
The combination of these two methods, has been recently proposed and reached a important success in single-band Hubbard-Anderson model \cite{CPADMFTnonequilibrium2022, CPADMFTnonequilibrium2023,CPADMFTDFT2016}.
However, the DMFT-CPA method developed aims only to the on-site disorder cases. The effects of off-diagonal disorder are substantial, particularly when the energy levels of the host and impurity atoms exhibit a large bandwidth discrepancy. Under these conditions, the CPA becomes less effective, necessitating the development of computational methods that can account for off-diagonal disorder. Consequently, advancing techniques to solve the Anderson-Hubbard model with off-diagonal disorder is of paramount importance.

Despite the growing interest, research on strongly correlated systems with non-local disorder remains limited \cite{offdiagonal1989, offdiagonal1997, offdiagonal2001, BEBDMFT2021, randomhopping1993, randomhopping1994}. Notable exceptions include the BEB-CPA+DMFT approach \cite{BEBDMFT2021} and the pioneering work by Dobrosavljević and Kotliar, who employed functional integrals for quantum averaging within the DMFT framework \cite{randomhopping1993, randomhopping1994}. The auxiliary coherent potential approximation(ACPA) method, originally developed for phonon systems, has been validated in several studies \cite{ACPAPhononband2019, ACPAPhonon2021, ACPAPhonon2019, ACPAPhonontransport2019, Weiqi2022, ACPAclustertheory}. It relies on the principle of 'finiteness' in off-diagonal disorder, introducing an extended local degree of freedom, referred to as the coupling space, to encapsulate the effects of off-diagonal disorder within an auxiliary medium. This approach allows for the equal treatment of both diagonal and off-diagonal disorder by applying CPA-like self-consistency within the auxiliary coupling space. Initially proposed for addressing disorder in vibrational systems \cite{ACPAPhonon2019, ACPAPhononband2019}, the method has since been extended to cluster theory \cite{ACPAclustertheory} and to the calculation of quantum transport properties in disordered nanostructures.\cite{ACPAPhonontransport2019,cui2024}.

In this work, we extend the ACPA method to electronic systems and combine it with DMFT to provide an effective computational approach for disordered strongly correlated systems. We employ the hybridization expansion continuous-time quantum Monte Carlo (CT-QMC) method \cite{CTHYB2006PRL} as the DMFT impurity solver within the ACPA+DMFT framework. We apply this method to solve the Anderson-Hubbard model with off-diagonal disorder, calculating the Matsubara frequency Green's function and deriving the spectral function via analytical continuation. Our investigation focuses on the metal-insulator transition in strongly correlated systems subjected to both diagonal and off-diagonal disorder at finite temperatures. Notably, we observe a reentrance effect in the simple cubic lattice as the interaction strength $U$ increases. These findings are consistent with results from previous studies using CPA+DMFT \cite{Lombardo2006} and BEB-CPA+DMFT \cite{BEBDMFT2021}.

The organization of this article is as follows: Section \ref{II} covers the methodology, including an introduction to the ACPA method for electronic systems, its combination with the DMFT method to solve the Anderson-Hubbard model with off-diagonal disorders. 
Section \ref{III} presents the calculation results and discussions. 
Finally, we summarrize our work in Sec. \ref{IV} and provide more information in Append.\ref{AppendFourierTrans}. 

\section{Theory} \label{II}
In this section, we give a detailed derivation of ACPA formalism that applied to electronic system and its combination with DMFT.

\subsection{ACPA formalism for electronic systems}\label{ACPA}

We start with the Hamiltonian of an electron system, which contains both on-site and off-site disorders
\begin{equation}
\hat{H}=\sum_{ i,j ,\sigma} t^{Q_i,Q_j}_{i j} \hat{c}_{i \sigma}^{\dagger} \hat{c}_{j \sigma}+\sum_{i\sigma} (\epsilon^{Q_i}_{i}-\mu)\hat{n}_{i\sigma},
\label{HE}
\end{equation}
where $t_{i,j}^{Q_i,Q_j}$ is the hopping integral between site $i$, occupied by an atom $Q_i$, and site $j$, occupied by an atom $Q_j$,
$\epsilon_i^{Q_i}$ denotes the onsite energy at site $i$ when it is occupied by $Q_i$,
$\hat{c}_{i\sigma}^\dagger$ ( $\hat{c}_{j\sigma}$ ) is the creation (annihilation) operator, and $\hat{n}_{i\sigma} = \hat{c}_{i\sigma}^\dagger \hat{c}_{i\sigma}$ is the number operator,
$\mu$ is the chemical potential of the system. 

In the ACPA method, the central idea is to decompose the random hopping amplitudes in \myref{HE} into
\begin{equation}
t_{i  j }^{Q_i, Q_j}=x_{i}^{Q_i} S_{i  j } x_{j}^{Q_j}+\lambda_{ij}, 
\label{ACPAhopping}
\end{equation}
where $S$ and $\lambda$  
are matrices that independent of the chemical occupations on sites $i$ and $j$, and $x_i^{Q_i}$ and $x_j^{Q_j}$ generically should be fitted from the chemical dependent hopping parameters.

To proceed, we divided both side of the above equation by $x_i^{Q_i}$, yielding 
\begin{equation}
\frac{t_{i  j }^{Q_i, Q_j}}{x_i^{Q_i}}= S_{i  j } x_{j}^{Q_j}+\frac{\lambda_{i  j }}{x_i^{Q_i}}.
\label{ACPAhoppingtransfer}
\end{equation}
which is a linear combination of quantities of single-site occupation on the right-hand side.
For the given Hamiltonian $H$, all hopping terms are treated and can be expressed in the following form:
\begin{equation}
H=\mathcal{H}X,
\end{equation}
where $X$ is a random diagonal matrix with $X_{ii} = x_i^{Q_i}$, and $\mathcal{H}$ describes the auxiliary system. From \myref{ACPAhoppingtransfer}, it is evident that $\mathcal{H}$ can be expressed as the sum of single-site terms, namely,
\begin{equation}
\mathcal{H}=\sum_i\tilde{\mathcal{H}}^{i,Q_i},
\end{equation}
where we isolate the elements related to the occupation $Q_i$ of the $i$ site and collect them into $\tilde{\mathcal{H}}^{i,Q_i}$. For a specific occupation $Q_i$, the single-site matrix $\widetilde{\mathcal{H}}^{i,Q_i}$ for a one-dimensional single-band system with nearest-neighbour hopping can be given by,
\begin{equation}
\widetilde{\mathcal{H}}^{i,Q_i}=\begin{pmatrix}
  0 &S_{(i-1)  i  }x^{Q_i}_i & 0   \\
\\
  \frac{\lambda_{i  (i-1) }}{x^{Q_i}_i} &\frac{\epsilon^{Q_i}_i-\mu }{x^{Q_i}_i} & \frac{\lambda_{i  (i+1) }}{x^{Q_i}_i}  \\
\\
 0  &S_{(i+1)  i }x^{Q_i}_i  &0
\end{pmatrix}.
\label{KmatrixACPA}
\end{equation}
\begin{figure}[t]
\centering
\includegraphics[width=0.9\linewidth]{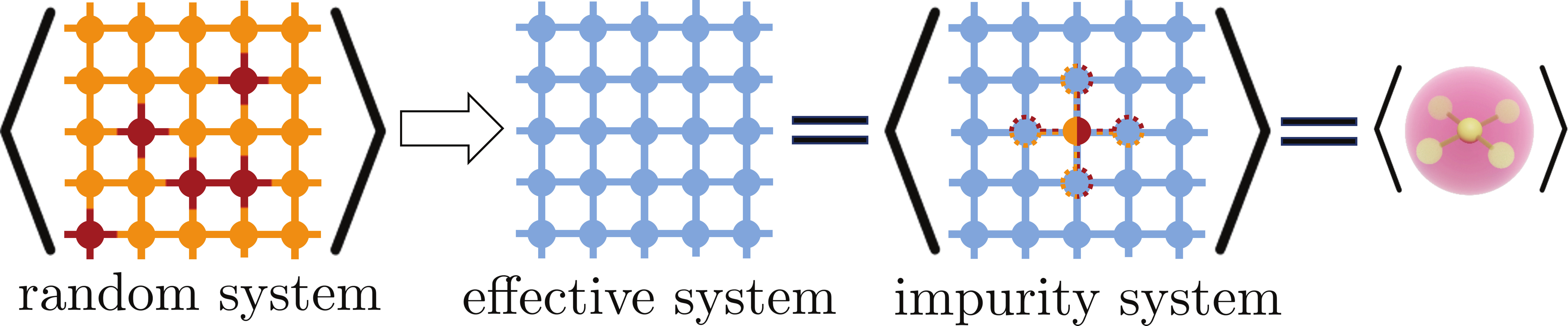}
\caption{Schematic illustration of ACPA idea in a 2D lattice model.  The orange (brown) color represents species A (B), while the blue represents the averaged effective medium. Spheres denote the diagonal components, and rods represent the off-diagonal components. $\langle \cdots \rangle$ denotes the disorder average over the configurational space.}
\label{ACPALOOP}
\end{figure}
For each site with $Z$ neighboring sites and $N$ orbital, the dimension of $\widetilde{\mathcal{H}}^{i,Q_i}$  is $N(Z+1)$. We introduce a coupling space $\mathcal{C}$ to encapsulate the extended degree of freedom in the single-site $\widetilde{\mathcal{H}}^{i,Q_i}$ \cite{ACPAclustertheory}, which is beyond the conventional single-site dimension N. This coupling space is expressed as the direct product $\mathcal{C} = \mathcal{S} \otimes \mathcal{T}$ in which $\mathcal{S}$ denotes the single-site orbital space, and $\mathcal{T}$ is formed by orthonormal bases $\{ |\mathbf{T}_{j^{\prime}}\rangle, (j^{\prime}=0,1,...,N) \}$, where $\mathbf{T}_0 = 0$ and $\mathbf{T}_{j^{\prime}} (j^{\prime}=1,...,Z)$ denotes the translational vectors to the neighboring sites of site $i$, namely $\mathbf{R}_{i+j^{\prime}}=\mathbf{R}_i+\mathbf{T}_{j^{\prime}}$, with the hopping element $t_{(i+j')i} \ne 0$.
We can rewrite $\widetilde{\mathcal{H}}^{i,Q_i}$ in coupling space $\mathcal{C}$, as follows,
\begin{equation}
\begin{aligned}
\widetilde{\mathcal{H}}^{i,\mathcal{C},Q_i}=&\sum_{j^{\prime}\neq0}{ {S_{(i+j^{\prime})i}x_i^{Q_i}|\mathbf{T}_{j^{\prime}}}\rangle\langle{\mathbf{T}_0}|}+\sum_{j^{\prime}\neq0}{\frac{\lambda_{i(i+j^{\prime})}}{x_i^{Q_i}}|\mathbf{T}_0\rangle\langle{\mathbf{T}_{j^{\prime}}}|} \\
&+\frac{\epsilon_i^{Q_i}-\mu}{x_i^{Q_i}}|\mathbf{T}_0\rangle\langle{\mathbf{T}_0}|.
\end{aligned}
\end{equation}
In the following, the quantity with the superscript $\mathcal{C}$ is defined in the coupling space $\mathcal{C}$ with a dimension of $N(Z+1)$.
Based on the above definitions of auxiliary systems, for a specific atomic configuration, we can express the lattice Green's function(GF) $G$ with an auxiliary Green's function $g$, 
\begin{equation}
G=(z-H)^{-1}=(z-\mathcal{H}X)^{-1}=gX^{-1},
\end{equation}
where $g=P^{-1}$ and $P=(zX^{-1}-\mathcal{H})=\sum_{i}\widetilde{P}^{i,\mathcal{C},Q_i}$. The single-site $P^{i,\mathcal{C},Q_i}$ is given as
\begin{equation}
\widetilde{P}^{i,\mathcal{C},Q_i}_{j,k}=\frac{z\delta_{j,i}\delta_{k,i}}{x^{Q_i}_i}-\widetilde{\mathcal{H}}^{i,\mathcal{C},Q_i}_{j,k},
\label{Pij}
\end{equation}
where $\delta_{i,j}$ is Dirac function. 
To do the disorder average, similarly to conventional CPA, we can introduce an effective medium  $\mathcal{P}$ whose auxiliary Green's function equals to the disorder averaged $\bar{g}$, namely
\begin{equation}
 \mathcal{\bar{g}} =\mathcal{P}^{-1},
\label{ag}
\end{equation}
and
\begin{equation}
 \mathcal{P} = \sum_i \mathcal{P}^{i,\mathcal{C}}
\end{equation}
where the single-site quantity $\mathcal{P}^{i,\mathcal{C}}$ represents the effective auxiliary Hamitonian defined in the coupling space. To form a self-consistent calculation of $\mathcal{P}$ , one can define an auxiliary coherent interactor $\Omega^{i,\mathcal{C}}$ to describe the average interaction of site $i$ with surrounding medium in $\mathcal{C}$, so that the signle-site $\mathcal{\mathcal{\bar{g}}}^{i,\mathcal{C}}=\left[\mathcal{P}^{i,\mathcal{C}} -\Omega^{i,\mathcal{C}} \right]^{-1}
\label{gi}$(in the single-site approximation), and then
\begin{equation}
\Omega^{i,\mathcal{C}}=\mathcal{P}^{i,\mathcal{C}}-\left[\mathcal{\mathcal{\bar{g}}}^{i,\mathcal{C}}\right]^{-1}.
\label{gi1}
\end{equation}
With $\Omega^{i,\mathcal{C}}$,  for the specific occupation $Q_i$, we have the single-site conditionally averaged $\widetilde{g}^{i,\mathcal{C},Q_i}$ as follows,
\begin{equation}
 \widetilde{\mathcal{g}}^{i,\mathcal{C},Q_i}=\left[ \widetilde{P}^{i,\mathcal{C},Q_i}-\Omega^{i,\mathcal{C}} \right]^{-1},
\label{gQ}
\end{equation}
Then, $\mathcal{P}^{i,\mathcal{C}}$ can be updated as
\begin{equation}
\mathcal{P}^{i,\mathcal{C}} =\left[\sum_{Q_i} c_i^{Q_i} (\widetilde{P}^{i,\mathcal{C},Q_i} -\Omega^{i,\mathcal{C}})^{-1}\right]^{-1}+\Omega^{i,\mathcal{C}}.
\label{AP}
\end{equation}
Eqs.(\ref{ag}-\ref{AP}) form a closed self-consistent ACPA loop for solving the disorder averaged auxiliary GF, to obtain the physical properties.  At the self-consistency, 
and the average  Green's funcition of $i$ site in coupling space is given by,
\begin{equation}
 \bar{g}^{i,\mathcal{C}}= \sum_{Q_i} c_{i}^{Q_i} \widetilde{g}^{i,\mathcal{C},Q_i}.
\end{equation}
 In the practical implementation, since the effective medium $\mathcal{P}$ is translational invariant, we use the Fourier transformation to the reciprocal space,  please refer to Append.\ref{AppendFourierTrans}. After the ACPA iteration in the auxiliary coupling space,
the relationship between the auxiliary and the site-diagonal element of the lattice GFs are given by,
\begin{equation}
G^{Q_i}_{ii}=(\widetilde{g}^{i,\mathcal{C},Q_i})_{ii}(x_i^{Q_i})^{-1}.
\label{phyG}
\end{equation}
and then
\begin{equation}
\bar{G}_{ii}= \sum_{Q_i} c_i^{Q_i}G^{Q_i}_{ii}.
\label{phyG1}
\end{equation}
As a result, we can perform the self-consistent ACPA in the coupling space and then convert the quantities to the physical space. The ACPA method transforms the off-diagonal disorder in the lattice space into auxiliary diagonal-like disorder in the coupling space, in which ACPA solution for the disorder averaged single-particle GF can be obtained self-consistently. 
However, the DMFT achieves a self-consistent solution directly in the lattice space for strongly correlated system.
Therefore, the key step to combine ACPA and DMFT for addressing disordered interacting systems lies in the bi-directive transformation between the auxiliary coupling space and physical space.
In the following, we outline the derivation process of ACPA+DMFT.

\subsection{DMFT for interacting systems}\label{DMFT}
\begin{figure*}
\centering
\includegraphics[scale=0.15]{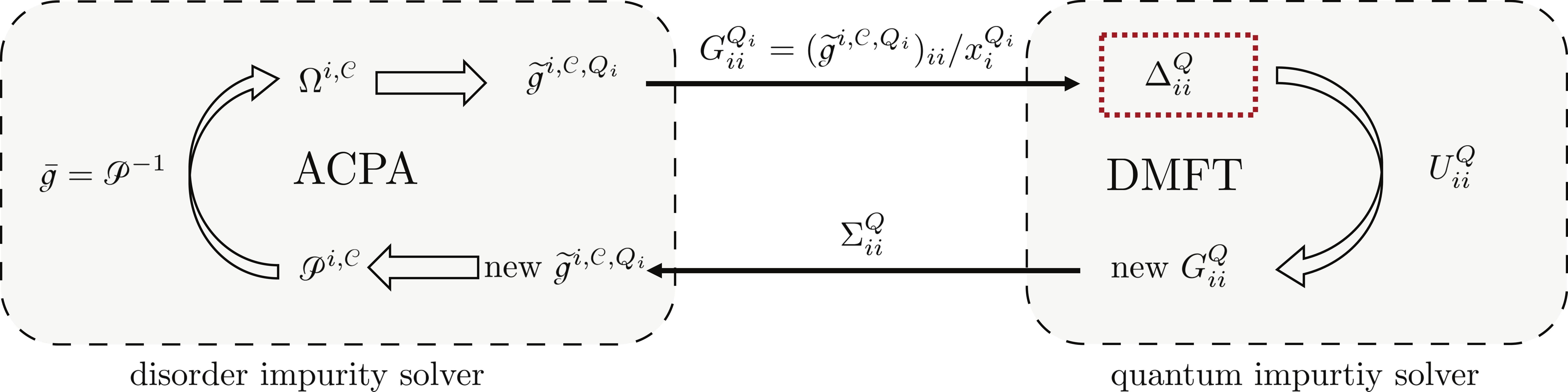}
\caption{Self-consistent loop of ACPA+DMFT for strongely correlated  Anderson-Hubbard model with off-diagonal disorders.}
\label{self-loop}
\end{figure*}

To begin, we consider the Hamiltonian of Anderson-Hubbard model for a random lattice configuration as follows
\begin{equation} 
\hat{H}=\hat{H}_0 +\sum_{i} U^{Q_i}_{i} \hat{n}_{i \uparrow} \hat{n}_{i \downarrow}, 
\label{A-H Hamilton}
\end{equation}
where $\hat{H}_0$ refers to the noninteracting Hamiltonian in \myref{HE}, and $U_i^{Q_i}$ is the on-site Hubbard interaction for $Q_i$ occupying the $i$ site. The theoretical goal is to obtain the disorder averaged single-particle GF for understanding various physical properties of the strongly corelated and disordered systems. Then, for a specific atomic configuration, the partition function of the system is $Z = \int \mathcal{D}[a^{\ast}, a]e^{-S}$, and the single-particle GF is given as \cite{Vollhardt2012}
\begin{equation}
G_{ij,\sigma}(\tau_1-\tau_2)= \frac{1}{Z}\int \mathcal{D}[a^{\ast}, a]a_{i,\sigma}(\tau_1) a_{j,\sigma}^{\ast}(\tau_2) e^{-S},\label{singleparGF}
\end{equation}
where  $a$/$a^{\ast}$ are Grassmann variables and the action $S$ of the system is 
given as
\begin{equation}
\begin{aligned}
S=\int_0^{\beta} &d\tau\Bigg\{ \sum_{i, j,\sigma}a^{\ast}_{i\sigma}(\tau)\Bigg[(\frac{\partial}{\partial\tau}+\epsilon^{Q_i}_{i}-\mu)\delta_{ij}+t^{Q_i,Q_j}_{i j}\Bigg ]\\ 
& a_{j \sigma}(\tau)
+\sum_{i} U^{Q_i}_{i}a^{\ast}_{i\uparrow}(\tau)a_{i \uparrow}(\tau)a^{\ast}_{i\downarrow}(\tau)a_{i \downarrow}(\tau)\Bigg\},
\end{aligned}
\end{equation}
where $\tau$ is imaginary time, and $\beta$ denotes the system's temperature.
To describe the single-particle GF in \myref{singleparGF}, one can represent the Green's function by introducing an effective single-particle Hamiltonian,
\begin{equation}
\hat{H}_{eff}=\hat{H}_0 +\hat{\Sigma},\label{Heffective}
\end{equation}
where  $\hat{\Sigma}$ is the self-energy that accounts for the strong-correlation effects of many-body interacting lattices,5 such that (in the energy domain)
\begin{equation}
G=\frac{1}{z-\hat{H}_0-\hat{\Sigma}}.
\label{sumGF}
\end{equation}
However, exact solution of $\Sigma$ is not possible for a lattice problem. Here, we adopt the DMFT that approximates $\Sigma = \sum_i \Sigma_{ii}$ in which the nonlocal terms $\Sigma_{ij}=0$, presenting a local approximation.(This approximation becomes exact in the infinite-dimensional limit \cite{DMFT1989, Vollhardt2012}.) In DMFT, the central theme is mapping the quantum many-body lattice to an local impurity model  \cite{RMPGeorges1996, RMPKotliar2006, Hartmann1989}, and the local impurity solution for $\Sigma_{ii}$ can be obtained through the impurity solver. The action of this local impurity problem satisfies the following relation
\begin{equation}
\begin{aligned}
\frac{1}{Z_{imp,i}}e^{-S_{imp,i}[a^{\ast}_{i\sigma},a_{i\sigma}]}=\frac{1}{Z}\int \prod_{j \neq i ,\sigma}\mathcal{D}[a_{j\sigma}^{\ast},a_{j\sigma}]e^{-S},
\end{aligned}
\end{equation}
where the $Z_{imp,i}=\int \prod_{\sigma}\mathcal{D}[a^{\ast},a]e^{-S_{imp,i}}$. In the DMFT, the local impurity action $S^{Q_i}_{imp,i}$ for site $i$ with occupation $Q_i$ can be expressed as \cite{DMFTRMP1996},
\begin{equation}
\begin{aligned}
S_{imp,i}^{Q_i}=&\int_0^{\beta}\int_0^{\beta}\mathrm{d}\tau_1\mathrm{d}\tau_2\sum_{\sigma}a^{\ast}_{i\sigma}(\tau_1)\Delta^{Q_i}_{ii,\sigma}(\tau_1-\tau_2)a_{i\sigma}(\tau_2)\\
&+\int_0^{\beta}\mathrm{d}\tau \sum_{\sigma}a^{\ast}_{i\sigma}(\tau)(\frac{\partial}{\partial\tau}+\epsilon^{Q_i}_{i}-\mu)a_{i \sigma}(\tau)\\ 
&+\int_0^{\beta}\mathrm{d}\tau U^{Q_i}_{i}a^{\ast}_{i\uparrow}(\tau)a_{i \uparrow}(\tau)a^{\ast}_{i\downarrow}(\tau)a_{i \downarrow}(\tau),
\end{aligned}
\end{equation}
where
\begin{equation}
\Delta^{Q_i}_{ii,\sigma}=-\sum_{j,k\neq i}t^{\ast,Q_j,Q_i}_{ji}G^{(i)}_{jk\sigma}t^{\ast,Q_k,Q_i}_{ki},
\label{deltaG}
\end{equation}
where $t^{\ast}$ represents the rescaled hopping, and $G_{jk\sigma}^{(i)}$ represents the Green's function after removing the $i$ site. The $\Delta^{Q_i}_{ii,\sigma}$ denotes the hybridization function describing the influence of the bath on the site $i$, and it can be obtained using the on-site GF from Eq.\ref{sumGF}, namely $G^{Q_i}_{ii,\sigma}$,
\begin{equation}
\Delta^{Q_i}_{ii,\sigma}=z+\mu-\epsilon_i^{Q_i}-\Sigma^{Q_i}_{ii,\sigma}-[G^{Q_i}_{ii,\sigma}]^{-1}.
\label{dmftdelta}
\end{equation}
Then, with the hybridazation $\Delta^{Q_i}_{ii,\sigma}$, the local impurity Green's function can be calculated as, by the impurity slovers, such as CT-QMC \cite{RMPCTQMC2011}, exact diagonalization \cite{EDQimao1994,EDCaffarel1994,EDCapone2007}, numerical  renormalization group \cite{RMPNRG1975,RMPNRG2008},
\begin{equation}
G^{Q_i}_{imp,\sigma}=-\frac{1}{Z_{loc}}\int \prod_{\sigma}\mathcal{D}[a_{\sigma}^{\ast},a_{\sigma}]a_{i\sigma}a^{\ast}_{i\sigma}\mathrm{exp}[-S^{Q_i}_{imp,i}].
\label{Glocimpurity}
\end{equation}
The, the local self energy $\Sigma^{Q_i}_{ii,\sigma}$ is obtained as
\begin{equation}
\Sigma^{Q_i}_{ii,\sigma}=z +\mu -\Delta^{Q_i}_{ii,\sigma}-\epsilon_i^{Q_i}-[G^{Q_i}_{imp,\sigma}]^{-1}.
\label{dmftsigma}
\end{equation}
As a result, the lattice problem is reduced to a local impurity problem,  Eq.\ref{sumGF} and Eqs.(\ref{dmftdelta})-(\ref{dmftsigma}) form the closed set of DMFT self-consistent Equations for solving the single-particle GF. At the self-consistency, we have the following relation satistified, namely
\begin{equation}
G^{Q_i}_{ii,\sigma}=G^{Q_i}_{imp,\sigma}.
\end{equation}
\subsection{ACPA+DMFT formalism for disordered interacting systems}\label{ACPADMFT}
For the disordered Andersen-Hubbard model in Eq.\ref{A-H Hamilton}, we need to perform the disorder average to the single-particle GF in Eq.\ref{sumGF} to make it physically meaningful, and to restore the translational invariance to make the DMFT formalism practical. 
In the DMFT approximation, the $H_{eff}$ in Eq.\ref{Heffective} for a specific configuration can be explicitly written as (Here, we consider the case with SU(2) symmetry, so $\Sigma^{Q_i}_{ii}$ is independent of spin),
\begin{equation}
\hat{H}_{eff}=\sum_{i, j,\sigma} t^{Q_i,Q_j}_{i j} \hat{c}_{i \sigma}^{\dagger} \hat{c}_{j \sigma}+\sum_{i \sigma}\left(\epsilon^{Q_i}_{i}-\mu+\Sigma^{Q_i}_{ii}\right) \hat{n}_{i \sigma}.
\end{equation}
To apply ACPA for handling the off-diagonal disorder in above, we rewritte the auxiliary single-site quantity (in Eq.\ref{Pij}) in the coupling space to,
\begin{equation}
\widetilde{P}^{i,\mathcal{C},Q_i}_{j,k}=\frac{(z-\Sigma^{Q_i}_{ii})\delta_{j,i}\delta_{k,i}}{x^{Q_i}_i}-\widetilde{\mathcal{H}}^{i,\mathcal{C},Q_i}_{j,k}.
\label{Pijnew}
\end{equation}
which is key for implementing the ACPA depicted in Sec. \ref{ACPA}.
Then, the self-consistent loop of ACPA in combination with DMFT, for treating the disordered Anderson-Hubbard model, can be illustrated as follows, (also see \cref{self-loop}).\\
(1) Guess the initial interactor $\Omega^{i,\mathcal{C}}$;\\
(2) Calculate conditionally averaged auxiliary Green's function $\widetilde{\mathcal{g}}^{i,\mathcal{C},Q_i}$ with the given $\Omega^{i,\mathcal{C}}$ via \myref{gQ};\\
(3) Calculate lattice Green's function $G_{ii}^{Q_i}$ via \myref{phyG};\\
(4) Solve the hybridization function $\Delta_{ii}^{Q_i}$ in physical space via \myref{dmftdelta}; \\    
(5) Solve the impurity problem to obtain the self-energy $\Sigma_{ii}^{Q_i}$;\\
(6) Update $\widetilde{P}^{i,\mathcal{C},Q_i}$ and calculate the impurity Green's function $\widetilde{g}^{i,\mathcal{C},Q_i}$ in the coupling space via \myref{gQ};\\
(7) Determine $\mathcal{P}^{i,\mathcal{C}}$ by \myref{AP} to obtain $\mathcal{P}$, and calculate the average auxiliary Green's function $\mathcal{\bar{g}}$ by \myref{ag}, to update the auxiliary interactor $\Omega^{i,\mathcal{C}}$ via \myref{gi1} .\\
(8) Repeat steps (2)-(7) until $\Omega^{i,\mathcal{C}}$ and $\Delta_{ii}^{Q_i}$ converges.

After reaching the self-consistency, the averaged onsite Green's function $\bar{G}_{ii}$ in the physical space can be obtained using \myref{phyG} and \myref{phyG1}. 
It is worth noting that the hybridization function in the physical space depends on the occupied atom type $Q_i$, as a consequence of hopping disorder. In our present implementation of ACPA-DMFT, we  ultilize the hybridization expansion of CT-QMC method as implemented in Ref. \cite{Huang_2015}, and the real-frequency spectral functions is obtained using the maximum entropy method \cite{ME1991,ME1996} implemented in the ACFlow package \cite{HUANG2023}.

\section{Results and Discussions} \label{III}

\begin{figure}
\centering
\includegraphics[scale=0.2]{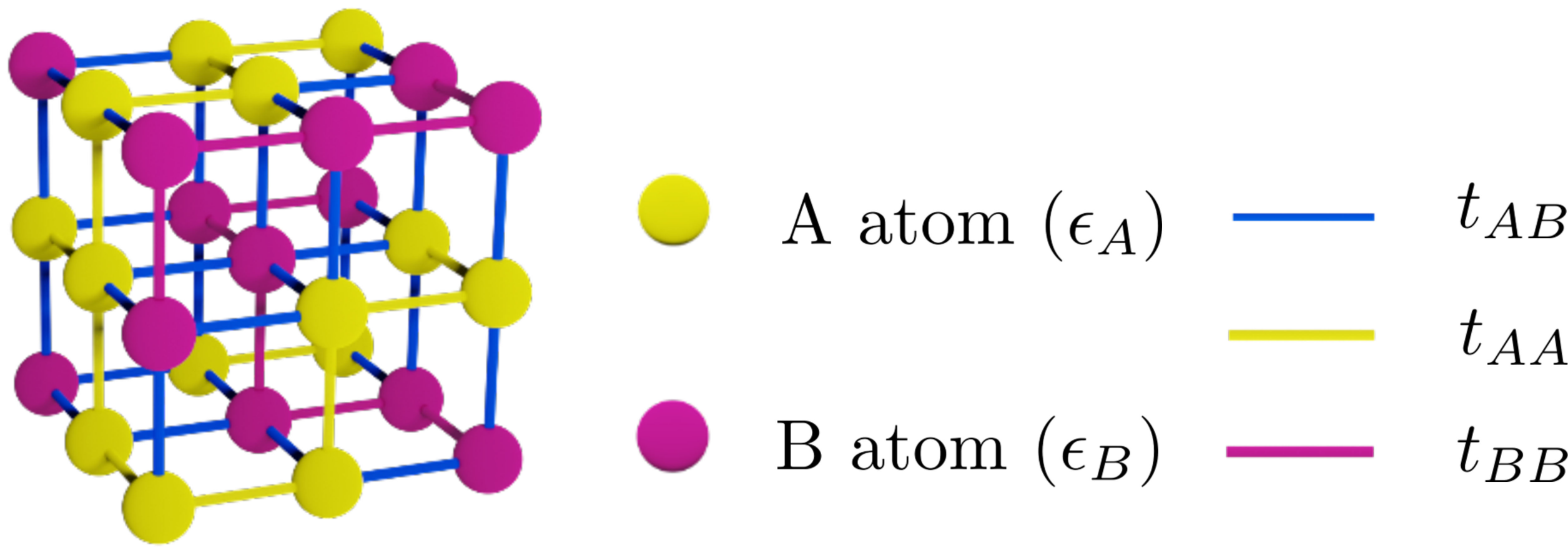}
\caption{Schematic illustration of a simple cubic lattice with both diagonal and off-diagonal disorders. }
\label{lattice}
\end{figure}

\subsection{Disordered non-interacting ($U=0$) limit}\label{Noninteractionlimit}
To demonstrate the applicability of ACPA method to disordered electronic systems, we compare our results with BEB-CPA and supercell (SPC) calculations for a noninteracting simple cubic lattice model with both diagonal and off-diagonal disorders (with the nearest-neighbour hopping) as shown in Fig.\ref{lattice}. The supercell calculations use the $5\times5\times5$ supercell size, and  $11\times11\times11$ k-mesh for BZ intergration. We calculate the alloy with the fixed $c_A=c_B=0.5$, and $t_{BB}=0.5$.  We use the energy broadening parameters $\eta = 0.05$ and $101\times101\times101$ k-mesh is used to ensure the convergence of BZ integration in ACPA and BEB-CPA calculations.   

\begin{figure}
\centering
\includegraphics[scale=0.31]{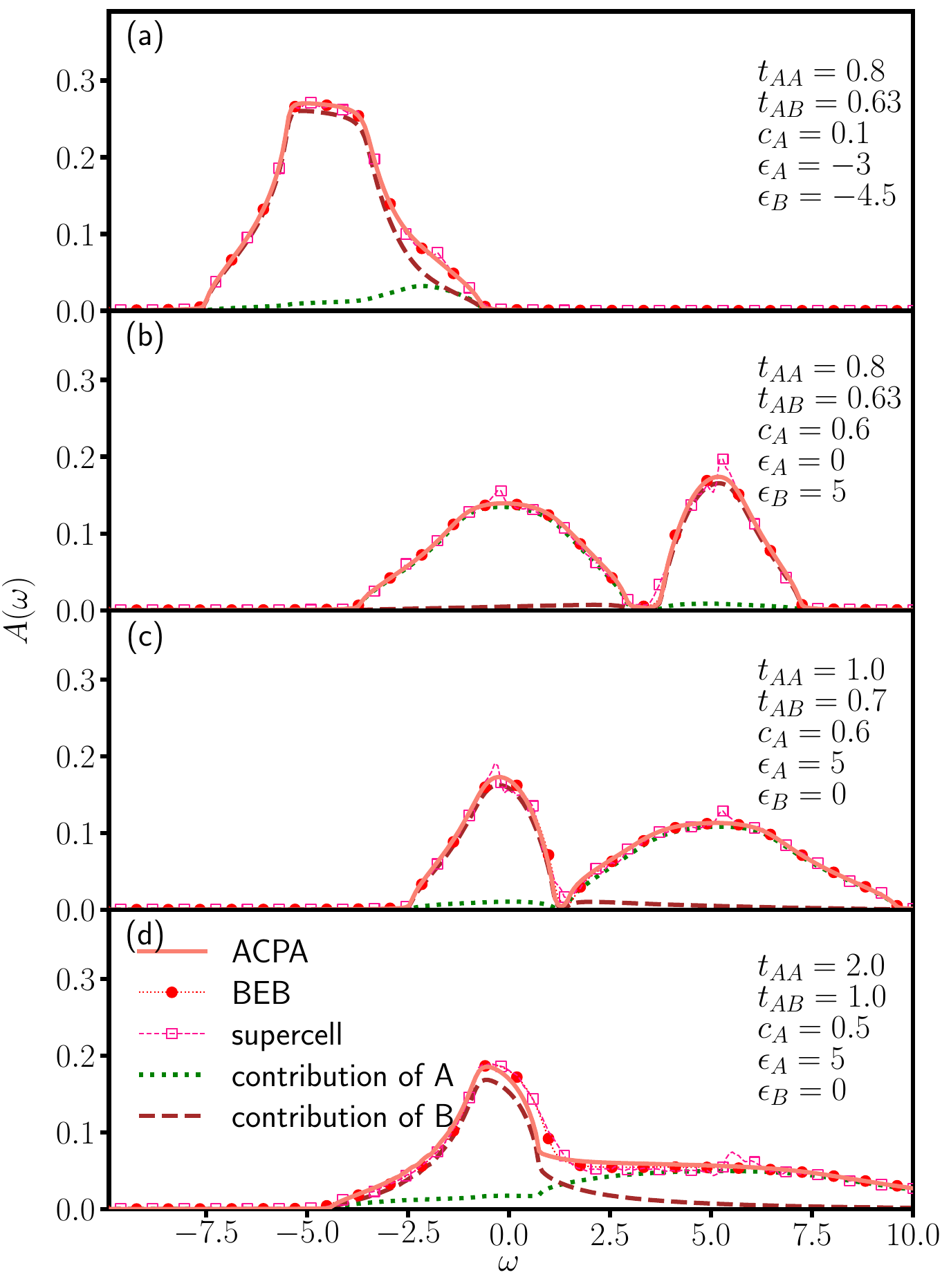}
\caption{A comparison of density of state (DOS) with the ACPA, BEB-CPA, and SPC methods in a simple cubic disordered lattice. The green dots and brown dashed lines represent the DOS contributions of species A and B for ACPA calculations, respectively. The solid orange line shows the total DOS for ACPA. The fixed parameters are  $t_{BB} = 0.5$, $c_A = c_B = 0.5$.}
\label{ACPAMATCHBEB}
\end{figure}

\cref{ACPAMATCHBEB} shows the averaged density of state (DOS) calculated by ACPA (solid line), BEB-CPA (sphere line) and SPC (square line) methods  for different parameters of $t_{AA}$, $t_{AB}$, $\epsilon_{A/B}$ as  provided in the legends of \cref{ACPAMATCHBEB}(a-c).
It is clear, for all the cases, ACPA results present very good agreement with the supercell and BEB-CPA calculations. In particular, as shown in \cref{ACPAMATCHBEB}, all methods present almost identical results, for example, in (a) for weak disorders in both $\epsilon$ and $t$, in (b) for weak disorder in $t$ and strong disorder in $\epsilon$, in (c) for moderate disorder strength in $t$ and strong $\epsilon$, and in (d) for strong disorders in $t$ and $\epsilon$ (except for the slight difference in $\omega \in [-1, 2]$). The minor deviation in \cref{ACPAMATCHBEB}(d) between ACPA and BEB-CPA and SPC results can be attributed to the facts that ACPA treat off-diagonal and diagonal disorders on the same footing within the coupling space, while BEB-CPA transforms the disorders into an diagonal disorder problem in the augmented space, presenting the difference between ACPA and BEB-CPA methods.
Compared to \cref{ACPAMATCHBEB}(a), as the diagonal disorder increases to the case (b), the DOS contributions of A and B atoms become separated, resulting in the system's DOS exhibiting a gap.
 By comparing with \cref{ACPAMATCHBEB}(b), as the off-diagonal disorder increases (by increasing $t_{AA}$ and $t_{AB}$),  DOSs of A and B atoms are gradually merged and the gap becomes eventually disappeared in the case (d). By changing the disorder strength in $t$, the DOS contribution of A is evidently broadened (resulting in the importnat reduction in the amplitude of DOS for $\omega \in [2.0,10.0]$), while DOS contribution of B presents important increase for $\omega \ge 0.5$, presenting the important effects of the $t$ disorders. In particular, at $\omega=5.0$, the DOS cotribution of A is 0.050 in (d), much smaller than 0.121 in (b) and 0.081 in (c), and at $\omega=0.5$ the DOS cotribution of A is 0.008 in (b) and 0.012 in (c), much smaller than 0.017 in (d). The agreement of ACPA with SPC and BEB-CPA results demonstrates the important effectiveness of the ACPA method in addressing disordered electronic systems, laying the ground for its extension to strongely correlated electronic systems with important atomic disorders. 

Concerning the computational cost of ACPA and BEB-CPA for the alloys, for an energy, computation of Green's function on k-mesh for the BZ integration generally dominate the ACPA and BEB-CPA self-consistent calculations. For each $\mathbf{k}$ point, the computational cost for ACPA in Eq.\ref{auxiliarygk} is $O(N^3)$, while the BEB-CPA cost is $O(M^3N^3)$ (M is the number of alloying species, N is the number of orbitals per primitive cell). This important difference between ACPA and BEB-CPA in the cost is attributed to the fact that the auxiliary $\bar{g}(\mathbf k)$ has the dimension of N (independent of the atomic species), whereas in BEB-CPA, the augmented GF (Fourier transformed) has the dimension of $MN$. As a result, ACPA holds the important promise for disorders with multiple orbitals, multiple components and Anderson-type off-diagonal disorders. 

\subsection{Disordered interacting cases} \label{ACPADMFTresults}
We next switch on the Coulomb interaction $U$ and study more realistic cases with both diagonal and off-diagonal disorders and strong correlations. The simple cubic lattice, as illustratred in Fig.\ref{lattice}, is investigated throughout this subsection as a model study to demonstrate the important effects of interplay between disorders and strong correlation.

\subsubsection{Influence of alloying concentration and $U$}

\begin{figure}[t]
\centering
\includegraphics[scale=0.31]{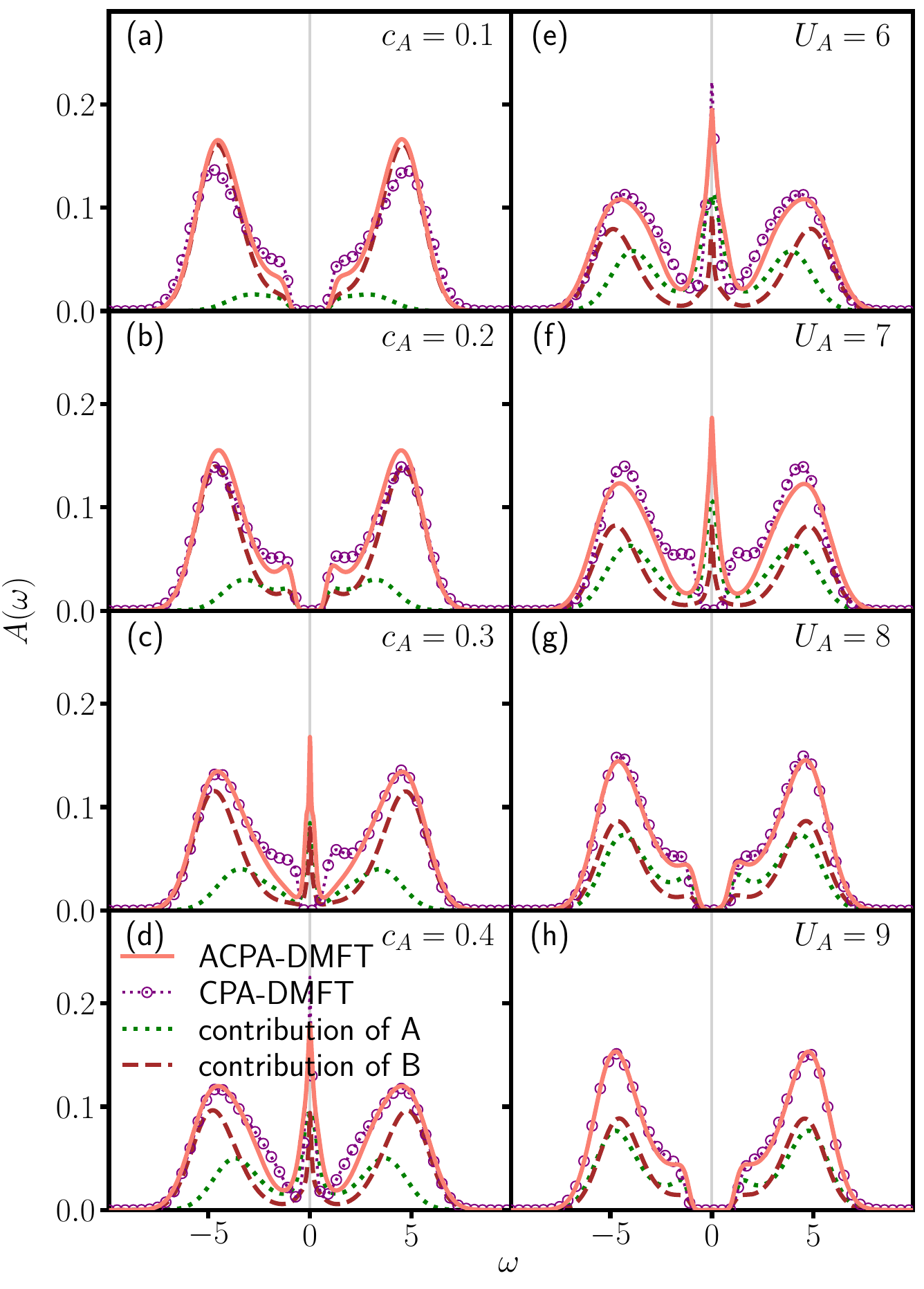}
\caption{ A comparison of DOS with the ACPA-DMFT (solid lines) and CPA-DMFT (circles) for a simple cubic lattice under different concentration of A (in (a-d)) and $U^A$ (in (e-h)). (a-d) use the fixed parameters $U_A = 6$, $U_B = 9$, and  
(e-h) use the fixed $U_B = 9$ and $c_A = 0.5$.
The hopping parameters for ACPA-DMFT are set as $t_{AA} = 0.8$, $t_{BB} = 0.5$, and $t_{AB} = 0.63$, while for CPA-DMFT, they are $t_{AA} = t_{AB} = t_{BB} = 0.63$. The onsite energy $\epsilon_A=-U_A/2,\epsilon_B=-U_B/2$ for satisfying the electron-hole symmetry for all the calculations}
\label{ACPAdifferentUAandCA}
\end{figure}

We first investigate the effects of doping concentration in the presence of off-diagonal disorders. We consider the fixed hopping disorders with the parameters $t_{AA} = 0.8$, $t_{BB} = 0.5$ and $t_{AB} = 0.63$, and fixed local disorders $U_A=6.0$, $U_B = 9$, $\epsilon_{A/B}=-U_{A/B}/2$ satisfying the half-filling condition. To compare with the ACPA-DMFT calculations with hopping disorder, we have aslo calculated the results without hopping disorder (using $0.63$ for all hoppings), denoted as  CPA-DMFT. \cref{ACPAdifferentUAandCA}(a-d) shows the DOS results for both ACPA-DMFT and CPA-DMFT calculations with different doping concentrations of A including $c_A=0.1$, $0.2$, $0.3$ and $0.4$. It is clear that all the DOS results satisfy the electron-hole symmetry, which is attributed to the half-filling simple-cubic lattce model. Importantly, as the doping concentration $c_A$ increases, the system undergoes a phase transition from metallic to insulating state (MIT) in both CPA-DMFT and ACPA-DMFT results, e.g. the the presence of quasiparticle peak in DOS at Fermi level ($\omega=0$). However, it is evident that the critical doping points differ between the two methods. With ACPA-DMFT including the disorders in the hopping $t$, the MIT occurs when $c_A$ lies between 0.2 and 0.3, while, with CPA-DMFT with constant $t$, the system still remains the metallic state at $c_A=0.3$ and changes to an insulating state at $c_A=0.4$. 
It should be mentioned that, for $c_A = 0$, the pure B lattice is a Mott insulator, while, for $c_A = 1.0$, the pure A lattice exhibits a metallic state. The observation, that a significant amount of dopant A, e.g. $c_A>0.2$, is required to present the MIT with ACPA-DMFT, suggest that the impact of alloying on the DOS cannot be simply viewed as some combination of the DOS from individual lattice of the two components. Instead, alloying alters the bath experienced by each of the component A and B atoms, thereby significantly changes the local DOS of A and B (see the DOS contribution of A and B, as shown) and influences the overall metal-insulator transition of the system. In particular, at $c_A=0.2$, there is a clear gap in DOS of A, while, at $c_A=0.4$, the DOS of B features evident metallic state at Fermi level, in contrast to the results of pure lattices of A and B. Furthermore, as seen in the post-transition regime (\cref{ACPAdifferentUAandCA}(c-d)) with ACPA-DMFT, the contribution of B atom to the quasiparticle peak remains substantial, presenting the important change of the bath of B by alloying with A. This study demonstrates that the phase transition can be effectively controlled via doping concentration, offering an effective approach for tuning the properties of strongly correlated system. It also highlights the important influence of off-diagonal disorder on the critical doping threshold, underscoring the necessity of considering the off-diagonal hopping disorders in such analyses.

\begin{figure}
\centering
\includegraphics[scale=0.31]{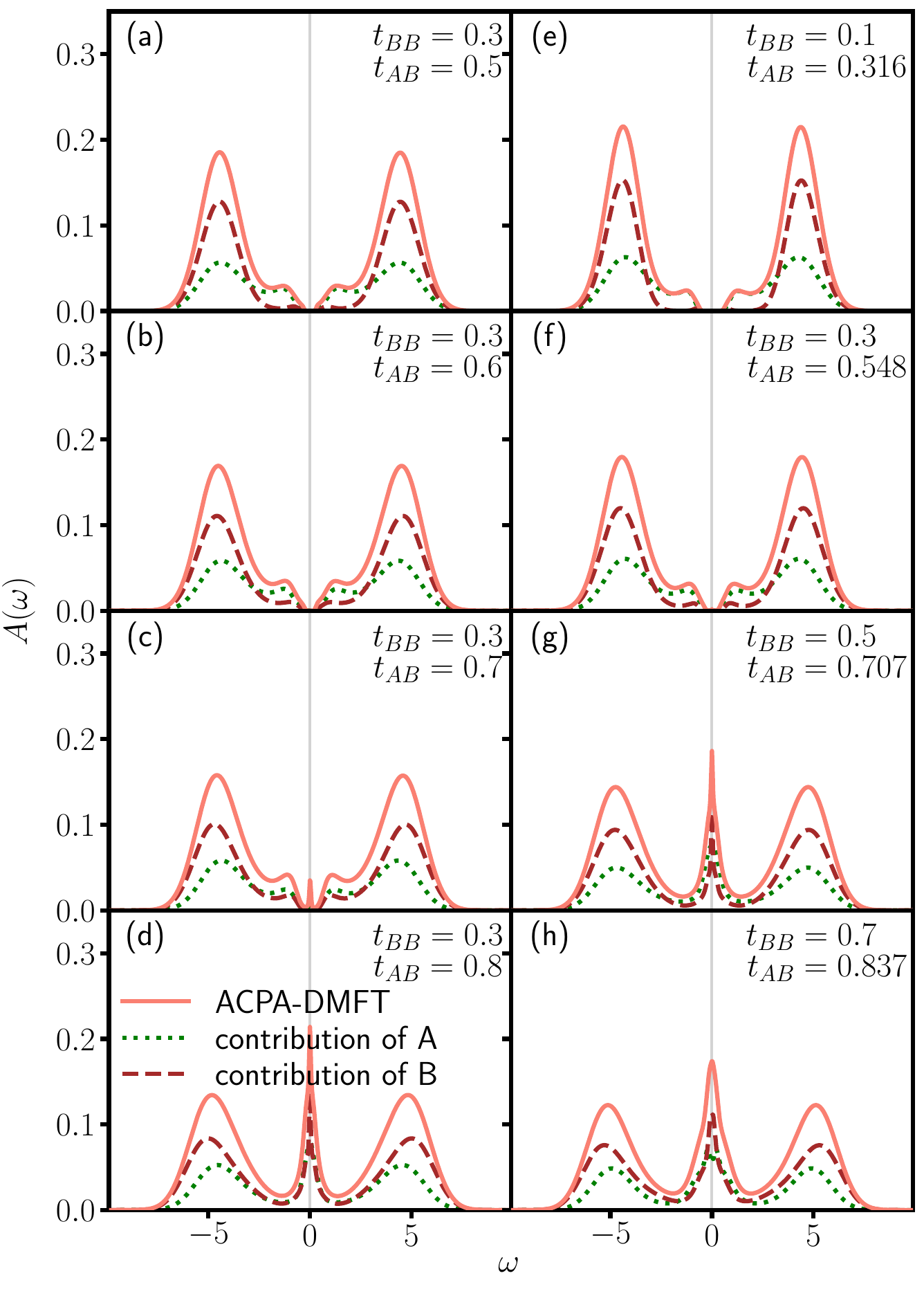}
\caption{DOS of ACPA-DMFT for a simple cubic lattice under varying hopping parameters of $t_{AB}$ in (a-d) and $t_{BB}$ in (e-h). (a-d) use the fixed $t_{AA}=1.0$ and $t_{BB} = 0.3$.  
(e-h) use the fixed $t_{AA}=1.0$ and $t_{AB} = \sqrt{t_{AA} t_{BB}}$. The model parameters $U_A = 8$, $U_B = 9$, $c_A = 1 - c_B = 0.4$ and $\epsilon_{A/B}=-U_{A/B}/2$ are used in all calculations.}
\label{ACPAdifferenttABandtBB}
\end{figure}

In the next, we invesigate the phase transition driven by varying the local disorders, including interaction $U_A$ and $\epsilon_A$, with disordered hopping by ACPA-DMFT and with constant hopping by CPA-DMFT (for comparison). For keeping the electron-hole symmetry in the DOS, we set $\epsilon_A=-U_A/2$ and $c_A=0.5$ with other system parameters same as \cref{ACPAdifferentUAandCA}(a–d). \cref{ACPAdifferentUAandCA}(e–h) shows the DOS results for $U_A=6.0$, $7.0$, $8.0$ and $9.0$ respectively. As illustrated  in \cref{ACPAdifferentUAandCA}(e-h), by increasing $U_A$, quasiparticle peak at the Fermi level gradually diminishes in both ACPA-DMFT and CPA-DMFT calculations. For example, ACPA-DMFT calculations presents a MIT between $U_A = 7$ and $U_A = 8$, while, for CPA-DMFT, the MIT occurs between $U_A = 6$ and $U_A = 7$, presenting the important influence of disordered hoppings. This indicates that the inclusion of hopping disorder can significantly modulate the interaction bath $\Delta^{Q_i}_{ii}$ for both A and B atoms. Especially for B, similar to \cref{ACPAdifferentUAandCA}(c–d), doping of component A induces the emergence of a quasiparticle peak in the DOS of B.  We can also find that, as increasing the $U_A$, the Hubbard bands associated with A become increasingly pronounced, ultimately the DOS contribution of A is comparable to that of B in the insulating states. As shown in \cref{ACPAdifferentUAandCA}(a, b) and (g, h), the quasiparticle peak from both A and B components simultaneously vanish or appear at the Fermi level, due to the implicit inter-dependence of the baths $\Delta^A$ and $\Delta^B$.

\subsubsection{Influence of hopping disorder}
\begin{figure}
  \centering  
  \subfigure{
     \label{quasiparticledifferenttAB}
     \centering
     \includegraphics[width=0.23\textwidth,height=0.22\textwidth]{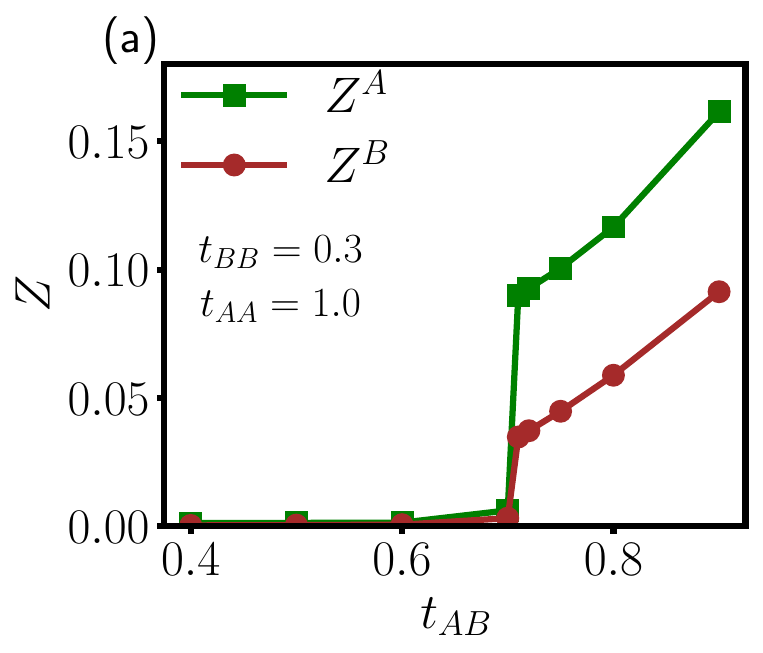}}
\hspace{-0.5cm}
\subfigure{
   \label{quasiparticledifferenttBB}
   \centering
   \includegraphics[width=0.23\textwidth,height=0.22\textwidth]{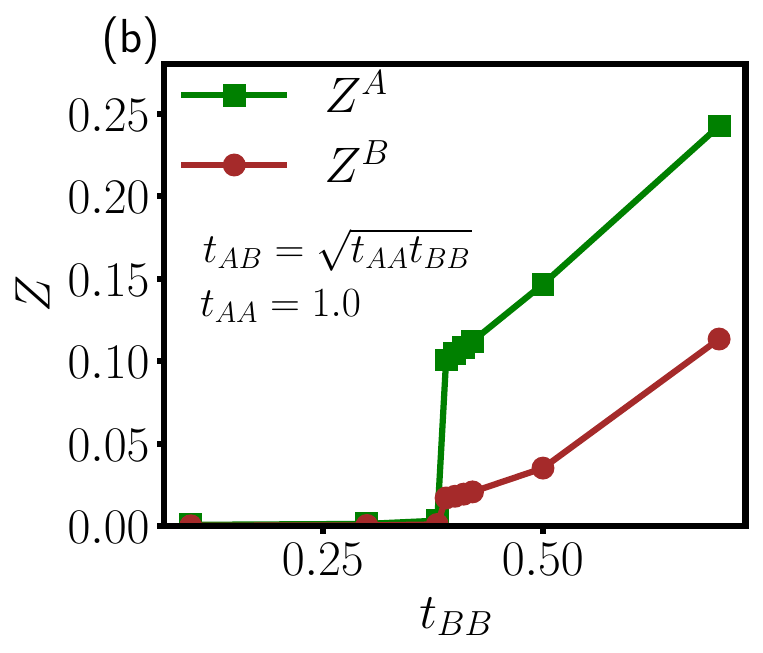}}
\caption{Quasi-particle weight $Z^{Q}$ (Q=A,B) versus  $t_{AB}$ in (a) and  $t_{BB}$ in (b). The fixed parameters $t_{AA} = 1.0$, $U_A = 8$, $U_B = 9$, $\epsilon_{A/B}=-U_{A/B}/2$ and  $c_A = 0.4$ are used in all calculations.}
\label{quasiparticle}
\end{figure}

\begin{figure}[b]
\centering
\includegraphics[scale=0.37]{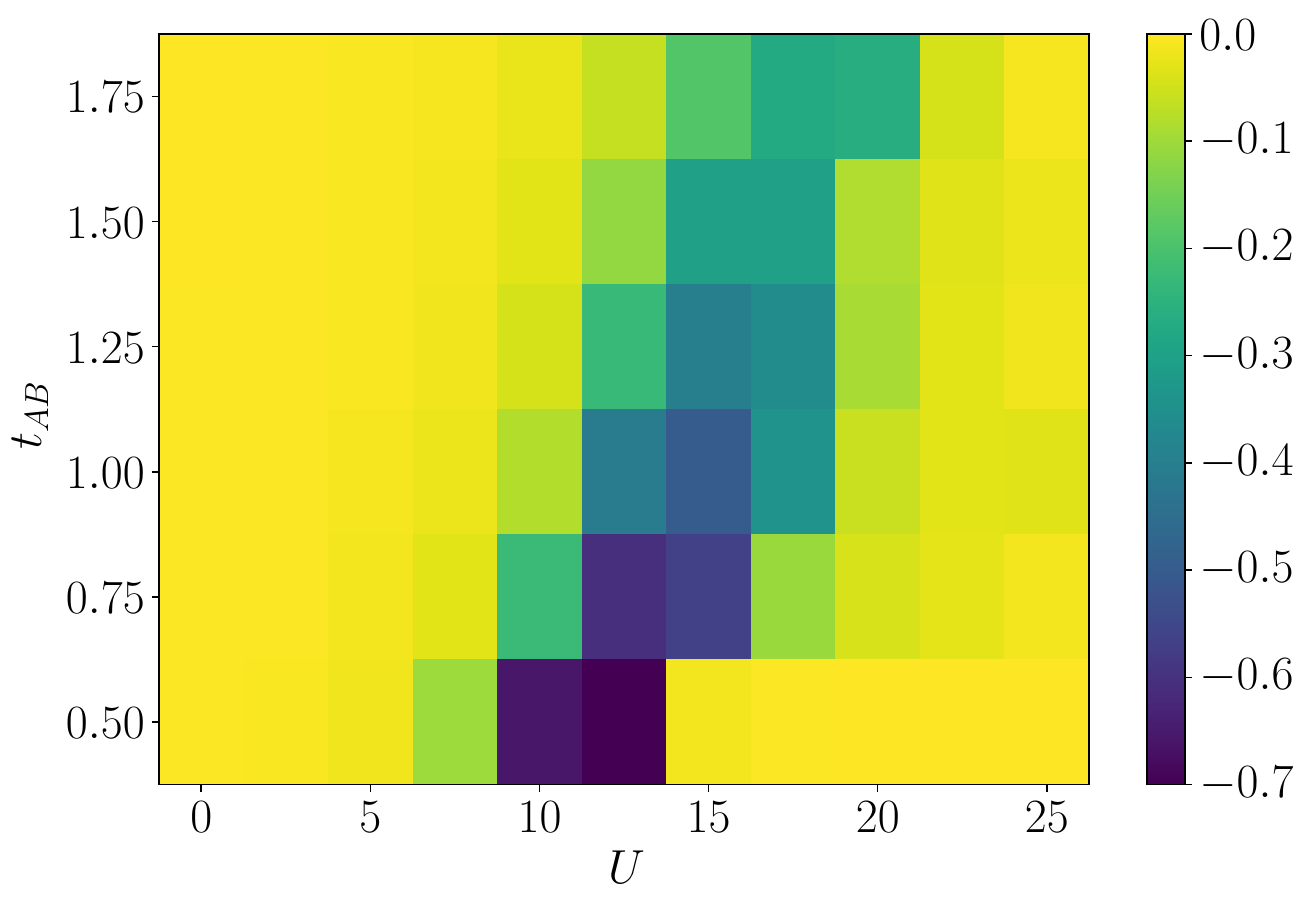}
\caption{Phase diagram of $\Im G(i\omega_0)$ in the $U$-$t_{AB}$, showing the imaginary part of the matsubara Green's function at $i\omega_0$. The fixed parameters $t_{AA}=t_{BB}=0.5$, $c_A=c_B=0.5$, $-(\epsilon_A + U_A/2) = (\epsilon_B + U_B/2) = 5$ and $U_A=U_B = U$ are used for all calculations.}
\label{fig:phase}
\end{figure}
To study the effect of hopping parameter $t_{AB}$ on the phase transition, we computed the DOS results for different $t_{AB}$ values by using ACPA+DMFT, as shown in \cref{ACPAdifferenttABandtBB}(a-d). We fix $t_{AA} = 1.0$, $t_{BB} = 0.3$, the interactions $U_A = 8$, $U_B =9$ and  $\epsilon_{A/B}=-U_{A/B}/2$. In \cref{ACPAdifferenttABandtBB}(a-d), we plot the DOS for $t_{AB}$ varying from $0.5$ to $0.8$. It is clear that, for both $t_{AB}=0.5$ in (a) and $0.6$ in (b), the systems show a evident gap in DOS, presenting the insulating states. However, as further increasing $t_{AB}$, the system undergoes the phase transition to metallic state. The increase of $t_{AB}$ can effectively increase the electronic kinetic energy to produce profound effect.
In \cref{ACPAdifferenttABandtBB}(c), we observe a very small and sharp peak in DOS at the Fermi level, which characterizes the vicinity of phase transition point near $t_{AB}=0.7$. We have checked that, by slightly increasing $t_{AB} = 0.71$, the quasiparticle peak at Fermi level ($\omega=0$) becomes much more pronouced than that in \cref{ACPAdifferenttABandtBB}(c), featuring the important characteristic of the Mott transition.
In \cref{ACPAdifferenttABandtBB}(d), both components A and B present the significant DOS contributions at the quasiparticle peak, illustrating that both components undergo a Mott transition. Therefore, it is seen that $t_{AB}$ has a significant influence on the state of the system, and can clearly drive a MIT. To further investigate the effect of the hopping parameter on the MIT, \cref{ACPAdifferenttABandtBB}(e-h) presnt the DOS results for different hopping $t_{BB}$ using ACPA-DMFT, with the fixed $t_{AA}=1.0$ and $t_{AB}=\sqrt{t_{AA}t_{BB}}$ (other system parameters are the same as those in \cref{ACPAdifferenttABandtBB}(a-d)). It is clear that changing the $t_{BB}$ can effectively drive a MIT in the binary alloy. The MIT presents for $t_{BB}$ between 0.3 and 0.5. By comparing the  \cref{ACPAdifferenttABandtBB}(c) and \cref{ACPAdifferenttABandtBB}(g) in which the $t_{AB}$ parameters are quite close, the change of $t_{BB}$ from 0.3 (in (c)) to 0.5 (in (g)) can dramatically enhance the quasi-particle peak. These observations underscore the importance of correctly handling off-diagonal disorder in theorectical simulation of disordered and strongly correlated systems.

To quantify the metal-insulator transition, we calculate the quasiparticle weights $Z^Q$ for the $Q$ component, defined as
\begin{equation}
Z^{Q} = \left[1 - \frac{\Im \Sigma^{Q}(i\omega_0)}{\omega_0}\right]^{-1},
\end{equation}
where $\Im \Sigma^{Q}(i\omega_0)$ is the imaginary part of the self-energy at the lowest Matsubara frequency $\omega_0$.
To analyze the phase transition, we investigate $Z^Q$ versus $t_{AB}$ and $t_{BB}$ in  the respective \cref{quasiparticle} (a) and (b). 
 As shown, with increasing  $t_{AB}$ or $t_{BB}$, quasiparticle weights for both A and B components exhibit a abrupt change from 0.0 to finite values, indicating a first-order phase transition. Such a discontinuity presents a clear hallmark of the Mott transition, consistent with previous discussions. In particular, the sharp change in $Z^Q$ happens at $t_{AB}=0.71$ in \cref{quasiparticle}(a), and at $t_{BB}=0.39$ in \cref{quasiparticle}(b), presenting the critical point for MIT. Notably, components A and B undergo the transition simultaneously, driven by the inter-component hopping  $t_{AB}$. 
 Furthermore, the quasiparticle weight of A is consistently higher than that of B due to the larger effective bandwidth of A (namely $t_{AA} > t_{BB}$). Beyond the transition point, the quasiparticle weights of both A and B grow nearly linearly with increasing $t_{AB}$ in \cref{quasiparticle}(a) and $t_{BB}$ in \cref{quasiparticle}(b). The observed MIT driven by varying $t_{AB}$ or $t_{BB}$ tells the fact that 
the phase transition is governed collectively by all hopping parameters, and underscores the important effects of disordered hopping interactions.

\begin{figure*}[htbp]
\centering
\includegraphics[scale=0.35]{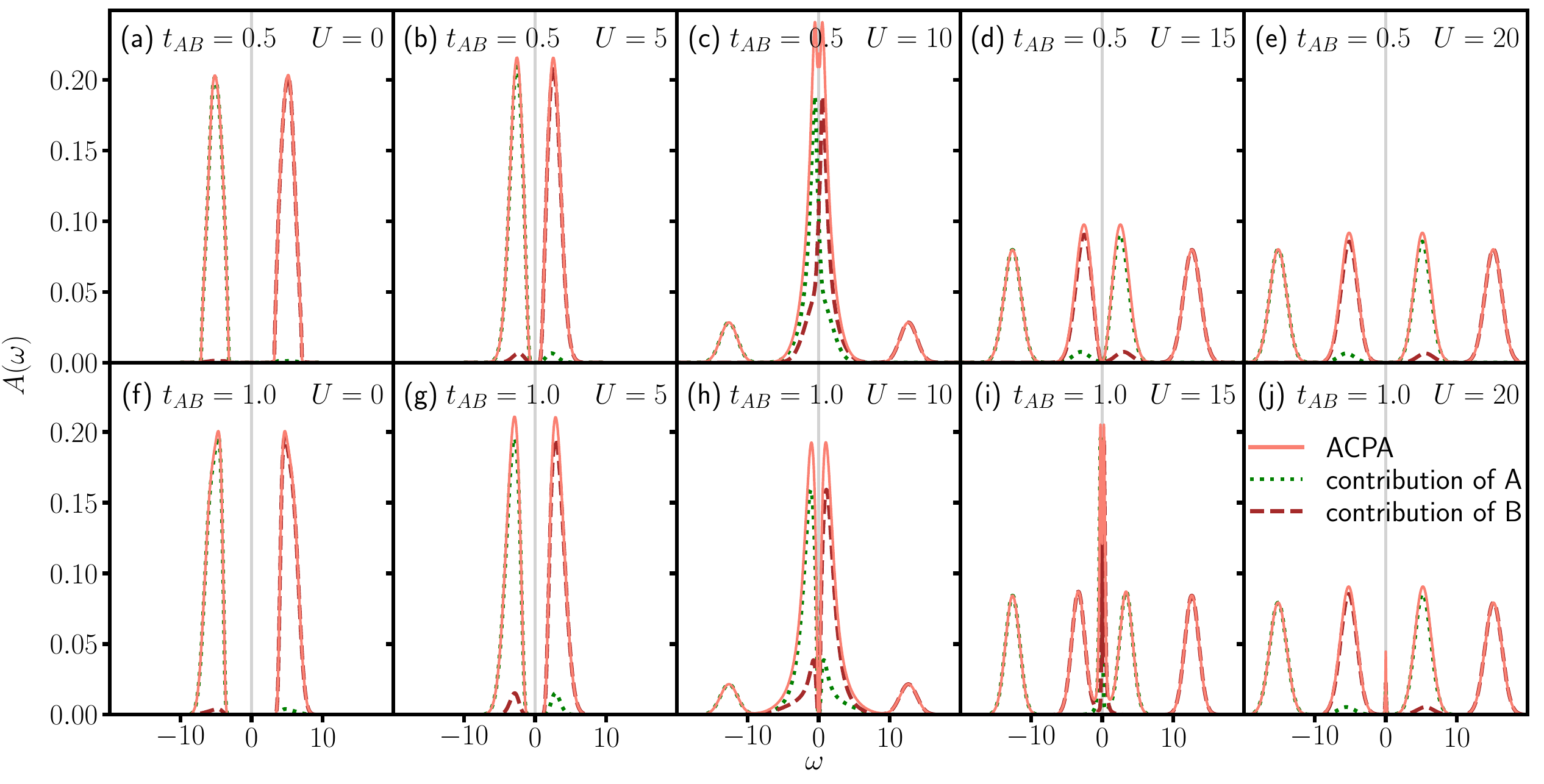}
\caption{Evolution of the DOS with  $U \in [0, 20]$ for $t_{AB} = 0.5$  (a-e,top row) and  $t_{AB} = 1.0$  (f-j,bottom row). The fixed parameters $t_{AA}=t_{BB}=0.5$, $c_A=c_B=0.5$, $-(\epsilon_A + U_A/2) = (\epsilon_B + U_B/2) = 5$, $U_A=U_B = U$ are used in all calculations.}
\label{nothalfdiagonaldifferentU}
\end{figure*}

\subsubsection{ Phase diagram for disordered and strongly correlated system}\label{reentrance}
For the reentrance phenomenon reported on the Bethe lattice in Ref.\cite{BEBDMFT2021}, we aim to observe a similar phenomenon on the simple cubic lattice. 
We next consider a general scenario where both onsite disorder and hopping disorder coexist, as commonly observed in real materials.
We focus on the case with $c_A = c_B =0.5$ and the fixed hopping amplitudes $t_{AA} = t_{BB} = 0.5$ (as Ref.\onlinecite{BEBDMFT2021}). The on-site energy and the Coulomb interaction are chosen $-(\epsilon_A + U_A/2) = (\epsilon_B + U_B/2) = 5$ and $U_A=U_B = U$, to ensure that we make the system half filled, but with species $A$ ($B$) electron (hole) doped. The crystal field splitting between $A$ and $B$ species in our calculation is $\delta = \epsilon_B - \epsilon_A = 10$. In \cref{fig:phase}, we plot the phase diagram of the disorder averaged $\Im G(i\omega_0)$ as a function of $U \in (0,25)$ and $t_{AB} \in (0.5, 1.75)$, where $\omega_0$ is the lowest Matsubara frequency point. The large negative value of $\Im G(i\omega_0)$ (dark blue) indicates the high DOS around the Fermi level, signifying a metallic phase.
Conversely, a small negative value (yellow) corresponds to a low DOS, therefore predicts the insulating states. As shown, for a fixed $t_{AB}$, increasing $U$ reveals a reentrance behavior: the system transitions from an insulating phase to an metallic phase and back to insulating again.
Additionally, as $t_{AB}$ increases, both $U_1$ and $U_2$ (the interaction strength where the system enter and leave the metallic states) also increases and the width of the metallic regime ($| U_2-U_1 |$) becomes larger as well. For example, at $t_{AB} = 0.5$, $| U_2 - U_1 | = 5$, and at $t_{AB} = 0.75$, $| U_2 - U_1 | = 7.5$.
This is because as $t_{AB}$ increases, the effective bandwidth of the system expands, reducing the ratio $U/D$ ($D$ is the effective bandwidth), which in turn slows down the phase transition, illustrating an important effect of disordered hoppings.

To explore the details further, we examine the spectral function for $t_{AB} = 0.5$ (for the CPA limit with only diagonal disorder, top row) and $t_{AB} = 1.0$ (bottom row) under varying interactions $U$ from $0$ to $20$, plotted from left to right in \cref{nothalfdiagonaldifferentU}. Evidently, As shown, with increasing $U$, the system transitions from an insulating state to a metallic state and then back to an insulating state, driven by the interaction U parameter. In particular, for both $t_{AB}$ cases, at $U = 0$, as shown in the \cref{nothalfdiagonaldifferentU}(a) and (f), the non-interacting spectrum exhibits band insulator behavior, where each band corresponds to one species of atoms. As increasing the $U=5$, the contributions of species $A$ and $B$ are shifted closer to the Fermi level, and the system still remains insulating in both $t_{AB}$ cases in \cref{nothalfdiagonaldifferentU}(b) and (g).
With further increase of $U$, shown in \cref{nothalfdiagonaldifferentU} (c) and (h), the spectral peaks for each species start to split into a satellite peaks due to the on-site interaction $U$.
The weight transfered to the sateillte peaks become increasingly prominent as $U$ further increases, as shown in \cref{nothalfdiagonaldifferentU} (d) and (i).
More importantly, the system is changed from insulating to metallic states at $U=10$ for $t_{AB}=0.5$ in \cref{nothalfdiagonaldifferentU}(d), and at $U=15$ for $t_{AB}=1.0$ in \cref{nothalfdiagonaldifferentU}(i). The further increase of U eliminate the quasi-particle peak at the Fermi level, changing the system back to the insulating state (e.g. at $U=15$ for $t_{AB}=0.5$ in \cref{nothalfdiagonaldifferentU}(d)).
As shown in \cref{nothalfdiagonaldifferentU}, It is clear that $t_{AB}=1.0$ (bottom row) can significantly delay phase transition from insulator to metal and from metal to insulator to larger U, compared to the results of $t_{AB}=0.5$ (top row), presenting the important effect of hopping disorders. In addition, at large U limit (e.g. $U = 20$ in \cref{nothalfdiagonaldifferentU} (e)), the results for species A and species B are similar to those obtained from the Hubbard-I approximation for each species, respectively \cite{HubbardI1963}. These results are consistent to BEB-CPA+DMFT calculations for the Bethe lattice \cite{BEBDMFT2021}.

\section{Summary} \label{IV}
We extended the ACPA to electronic systems and integrated it with DMFT, presenting an effective computational framework for investigating strongly correlated systems with both diagonal and off-diagonal disorder. Through a detailed study of a simple cubic lattice, we demonstrated the significant influence of off-diagonal disorder on MIT. Additionally, a reentrance phenomenon was observed, where the system transitions from insulating to metallic states and then back to insulating as the interaction strength increases. The results showed that off-diagonal disorder not only delays the onset of the transition but also broadens the range of the metallic phase, highlighting the intricate interplay between disorder and electron correlations.
The ACPA-DMFT method provides an effective approach for  for studying a wide range of disordered and strongly correlated systems, including both the equilibrium and nonequilibrium, \cite{ACPAPhonontransport2019,cui2024}, multi-orbital physics, and unconventional superconductivity. Furthermore, integrating ACPA-DMFT with first-principles calculations opens avenues for studying the properties of strongly correlated disordered alloys.

\section*{Acknowlegedgments}
Y.K. acknowledges financial support from NSFC (grant No. 12227901). J.Y. acknowledges the start-up funding provided by the Shanghai Institute of Microsystem and Information Technology. L.H. acknowledges financial support from NSFC (grant No. 12274380). The authors thank the HPC platform of ShanghaiTech University for providing the computational facility.
\appendix

\section{ACPA Fourier transform }\label{AppendFourierTrans}
In the practical implementation, since the effective medium $\mathcal{P}$ is translational invariant, with the Fourier transformation to the reciprocal space,we have
\begin{equation}
\begin{aligned}
\mathcal{P}(\mathbf{k})&=\sum_T{\mathcal{P}_{\mathbf{B},\mathbf{B+T}}\cdot e^{i\mathbf{k}\cdot\mathbf{T}}}
\end{aligned}
\end{equation}
where $\mathbf{B}$ denotes the basis in the primitive cell. $\mathcal{\bar{g}}(\mathbf{k})$ is computed using the following relations:
\begin{equation}
\mathcal{\bar{g}}(\mathbf{k})=\mathcal{P}(\mathbf{k})^{-1}.
\label{auxiliarygk}
\end{equation}
Then, the average auxiliary single-site Green's function in coupling space is given by
\begin{equation}
\mathcal{\bar{g}}^{i,\mathcal{C}}_{\mathbf{B}+\mathbf{T}_i,\mathbf{B}+\mathbf{T}_j}=\frac{1}{N_k}\sum_{\mathbf{k}}\mathcal{\bar{g}}(\mathbf{k})\cdot e^{-i\mathbf{k}\cdot(\mathbf{T}_i-\mathbf{T}_j)},
\end{equation} 
where $\mathbf{T}_{i/j}$ denote the translational vectors within the coupling space $\mathcal{C}$.

\bibliography{ACPADMFT}

\begin{thebibliography}{54}%
\makeatletter
\providecommand \@ifxundefined [1]{%
 \@ifx{#1\undefined}
}%
\providecommand \@ifnum [1]{%
 \ifnum #1\expandafter \@firstoftwo
 \else \expandafter \@secondoftwo
 \fi
}%
\providecommand \@ifx [1]{%
 \ifx #1\expandafter \@firstoftwo
 \else \expandafter \@secondoftwo
 \fi
}%
\providecommand \natexlab [1]{#1}%
\providecommand \enquote  [1]{``#1''}%
\providecommand \bibnamefont  [1]{#1}%
\providecommand \bibfnamefont [1]{#1}%
\providecommand \citenamefont [1]{#1}%
\providecommand \href@noop [0]{\@secondoftwo}%
\providecommand \href [0]{\begingroup \@sanitize@url \@href}%
\providecommand \@href[1]{\@@startlink{#1}\@@href}%
\providecommand \@@href[1]{\endgroup#1\@@endlink}%
\providecommand \@sanitize@url [0]{\catcode `\\12\catcode `\$12\catcode
  `\&12\catcode `\#12\catcode `\^12\catcode `\_12\catcode `\%12\relax}%
\providecommand \@@startlink[1]{}%
\providecommand \@@endlink[0]{}%
\providecommand \url  [0]{\begingroup\@sanitize@url \@url }%
\providecommand \@url [1]{\endgroup\@href {#1}{\urlprefix }}%
\providecommand \urlprefix  [0]{URL }%
\providecommand \Eprint [0]{\href }%
\providecommand \doibase [0]{https://doi.org/}%
\providecommand \selectlanguage [0]{\@gobble}%
\providecommand \bibinfo  [0]{\@secondoftwo}%
\providecommand \bibfield  [0]{\@secondoftwo}%
\providecommand \translation [1]{[#1]}%
\providecommand \BibitemOpen [0]{}%
\providecommand \bibitemStop [0]{}%
\providecommand \bibitemNoStop [0]{.\EOS\space}%
\providecommand \EOS [0]{\spacefactor3000\relax}%
\providecommand \BibitemShut  [1]{\csname bibitem#1\endcsname}%
\let\auto@bib@innerbib\@empty
\bibitem [{\citenamefont {Lee}\ and\ \citenamefont
  {Ramakrishnan}(1985)}]{Disorder1985}%
  \BibitemOpen
  \bibfield  {author} {\bibinfo {author} {\bibfnamefont {P.~A.}\ \bibnamefont
  {Lee}}\ and\ \bibinfo {author} {\bibfnamefont {T.~V.}\ \bibnamefont
  {Ramakrishnan}},\ }\href {https://doi.org/10.1103/RevModPhys.57.287}
  {\bibfield  {journal} {\bibinfo  {journal} {Rev. Mod. Phys.}\ }\textbf
  {\bibinfo {volume} {57}},\ \bibinfo {pages} {287} (\bibinfo {year}
  {1985})}\BibitemShut {NoStop}%
\bibitem [{\citenamefont {Evers}\ and\ \citenamefont
  {Mirlin}(2008)}]{AndersonTransitions2008}%
  \BibitemOpen
  \bibfield  {author} {\bibinfo {author} {\bibfnamefont {F.}~\bibnamefont
  {Evers}}\ and\ \bibinfo {author} {\bibfnamefont {A.~D.}\ \bibnamefont
  {Mirlin}},\ }\href {https://doi.org/10.1103/RevModPhys.80.1355} {\bibfield
  {journal} {\bibinfo  {journal} {Rev. Mod. Phys.}\ }\textbf {\bibinfo {volume}
  {80}},\ \bibinfo {pages} {1355} (\bibinfo {year} {2008})}\BibitemShut
  {NoStop}%
\bibitem [{\citenamefont {Imada}\ \emph {et~al.}(1998)\citenamefont {Imada},
  \citenamefont {Fujimori},\ and\ \citenamefont {Tokura}}]{MITRMP1998}%
  \BibitemOpen
  \bibfield  {author} {\bibinfo {author} {\bibfnamefont {M.}~\bibnamefont
  {Imada}}, \bibinfo {author} {\bibfnamefont {A.}~\bibnamefont {Fujimori}},\
  and\ \bibinfo {author} {\bibfnamefont {Y.}~\bibnamefont {Tokura}},\ }\href
  {https://doi.org/10.1103/RevModPhys.70.1039} {\bibfield  {journal} {\bibinfo
  {journal} {Rev. Mod. Phys.}\ }\textbf {\bibinfo {volume} {70}},\ \bibinfo
  {pages} {1039} (\bibinfo {year} {1998})}\BibitemShut {NoStop}%
\bibitem [{\citenamefont {Mott}(1990)}]{MITmott1990}%
  \BibitemOpen
  \bibfield  {author} {\bibinfo {author} {\bibfnamefont {N.}~\bibnamefont
  {Mott}},\ }\href {https://doi.org/10.1201/b12795} {\emph {\bibinfo {title}
  {Metal-Insulator Transitions (1st ed.)}}},\ Physical Sciences\ (\bibinfo
  {publisher} {CRC Press},\ \bibinfo {year} {1990})\BibitemShut {NoStop}%
\bibitem [{\citenamefont {Mott}(1949)}]{Mott_1949}%
  \BibitemOpen
  \bibfield  {author} {\bibinfo {author} {\bibfnamefont {N.~F.}\ \bibnamefont
  {Mott}},\ }\href {https://doi.org/10.1088/0370-1298/62/7/303} {\bibfield
  {journal} {\bibinfo  {journal} {Proceedings of the Physical Society. Section
  A}\ }\textbf {\bibinfo {volume} {62}},\ \bibinfo {pages} {416} (\bibinfo
  {year} {1949})}\BibitemShut {NoStop}%
\bibitem [{\citenamefont {Abrahams}(2010)}]{Andersonloc50}%
  \BibitemOpen
  \bibfield  {author} {\bibinfo {author} {\bibfnamefont {E.}~\bibnamefont
  {Abrahams}},\ }\href {https://doi.org/10.1142/7663} {\emph {\bibinfo {title}
  {50 Years of Anderson Localization}}}\ (\bibinfo  {publisher} {WORLD
  SCIENTIFIC},\ \bibinfo {year} {2010})\BibitemShut {NoStop}%
\bibitem [{\citenamefont {Zhu}\ and\ \citenamefont {Sheng}(2019)}]{ZhuPRL2019}%
  \BibitemOpen
  \bibfield  {author} {\bibinfo {author} {\bibfnamefont {W.}~\bibnamefont
  {Zhu}}\ and\ \bibinfo {author} {\bibfnamefont {D.~N.}\ \bibnamefont
  {Sheng}},\ }\href {https://doi.org/10.1103/PhysRevLett.123.056804} {\bibfield
   {journal} {\bibinfo  {journal} {Phys. Rev. Lett.}\ }\textbf {\bibinfo
  {volume} {123}},\ \bibinfo {pages} {056804} (\bibinfo {year}
  {2019})}\BibitemShut {NoStop}%
\bibitem [{\citenamefont {Prozorov}\ \emph {et~al.}(2024)\citenamefont
  {Prozorov}, \citenamefont {Sauls}, \citenamefont {Hirschfeld}, \citenamefont
  {Hussey},\ and\ \citenamefont {Iavarone}}]{disorderSC}%
  \BibitemOpen
  \bibfield  {author} {\bibinfo {author} {\bibfnamefont {R.}~\bibnamefont
  {Prozorov}}, \bibinfo {author} {\bibfnamefont {J.~A.}\ \bibnamefont {Sauls}},
  \bibinfo {author} {\bibfnamefont {P.}~\bibnamefont {Hirschfeld}}, \bibinfo
  {author} {\bibfnamefont {N.~E.}\ \bibnamefont {Hussey}},\ and\ \bibinfo
  {author} {\bibfnamefont {M.}~\bibnamefont {Iavarone}},\ }\bibfield  {journal}
  {\bibinfo  {journal} {Frontiers in Physics}\ }\textbf {\bibinfo {volume}
  {12}},\ \href {https://doi.org/10.3389/fphy.2024.1478445}
  {10.3389/fphy.2024.1478445} (\bibinfo {year} {2024})\BibitemShut {NoStop}%
\bibitem [{\citenamefont {Garg}\ \emph {et~al.}(2008)\citenamefont {Garg},
  \citenamefont {Randeria},\ and\ \citenamefont {Trivedi}}]{SPSC}%
  \BibitemOpen
  \bibfield  {author} {\bibinfo {author} {\bibfnamefont {A.}~\bibnamefont
  {Garg}}, \bibinfo {author} {\bibfnamefont {M.}~\bibnamefont {Randeria}},\
  and\ \bibinfo {author} {\bibfnamefont {N.}~\bibnamefont {Trivedi}},\ }\href
  {https://doi.org/10.1038/nphys1026} {\bibfield  {journal} {\bibinfo
  {journal} {Nature Phys.}\ }\textbf {\bibinfo {volume} {4}},\ \bibinfo {pages}
  {762} (\bibinfo {year} {2008})}\BibitemShut {NoStop}%
\bibitem [{\citenamefont {Chakraborty}\ \emph {et~al.}(2022)\citenamefont
  {Chakraborty}, \citenamefont {Löfwander}, \citenamefont {Fogelström},\ and\
  \citenamefont {Black-Schaffer}}]{SPSC2022}%
  \BibitemOpen
  \bibfield  {author} {\bibinfo {author} {\bibfnamefont {D.}~\bibnamefont
  {Chakraborty}}, \bibinfo {author} {\bibfnamefont {T.}~\bibnamefont
  {Löfwander}}, \bibinfo {author} {\bibfnamefont {M.}~\bibnamefont
  {Fogelström}},\ and\ \bibinfo {author} {\bibfnamefont {A.~M.}\ \bibnamefont
  {Black-Schaffer}},\ }\href {https://doi.org/10.1038/s41535-022-00450-w}
  {\bibfield  {journal} {\bibinfo  {journal} {npj Quantum Mater.}\ }\textbf
  {\bibinfo {volume} {7}},\ \bibinfo {pages} {44} (\bibinfo {year}
  {2022})}\BibitemShut {NoStop}%
\bibitem [{\citenamefont {Paalanen}\ \emph {et~al.}(1984)\citenamefont
  {Paalanen}, \citenamefont {Tsui}, \citenamefont {Gossard},\ and\
  \citenamefont {Hwang}}]{FQHE1984}%
  \BibitemOpen
  \bibfield  {author} {\bibinfo {author} {\bibfnamefont {M.}~\bibnamefont
  {Paalanen}}, \bibinfo {author} {\bibfnamefont {D.}~\bibnamefont {Tsui}},
  \bibinfo {author} {\bibfnamefont {A.}~\bibnamefont {Gossard}},\ and\ \bibinfo
  {author} {\bibfnamefont {J.}~\bibnamefont {Hwang}},\ }\href
  {https://doi.org/https://doi.org/10.1016/0038-1098(84)90343-0} {\bibfield
  {journal} {\bibinfo  {journal} {Solid State Communications}\ }\textbf
  {\bibinfo {volume} {50}},\ \bibinfo {pages} {841} (\bibinfo {year}
  {1984})}\BibitemShut {NoStop}%
\bibitem [{\citenamefont {Lombardo}\ \emph {et~al.}(2006)\citenamefont
  {Lombardo}, \citenamefont {Hayn},\ and\ \citenamefont
  {Japaridze}}]{Lombardo2006}%
  \BibitemOpen
  \bibfield  {author} {\bibinfo {author} {\bibfnamefont {P.}~\bibnamefont
  {Lombardo}}, \bibinfo {author} {\bibfnamefont {R.}~\bibnamefont {Hayn}},\
  and\ \bibinfo {author} {\bibfnamefont {G.~I.}\ \bibnamefont {Japaridze}},\
  }\href {https://doi.org/10.1103/PhysRevB.74.085116} {\bibfield  {journal}
  {\bibinfo  {journal} {Phys. Rev. B}\ }\textbf {\bibinfo {volume} {74}},\
  \bibinfo {pages} {085116} (\bibinfo {year} {2006})}\BibitemShut {NoStop}%
\bibitem [{\citenamefont {Metzner}\ and\ \citenamefont
  {Vollhardt}(1989)}]{DMFT1989}%
  \BibitemOpen
  \bibfield  {author} {\bibinfo {author} {\bibfnamefont {W.}~\bibnamefont
  {Metzner}}\ and\ \bibinfo {author} {\bibfnamefont {D.}~\bibnamefont
  {Vollhardt}},\ }\href {https://doi.org/10.1103/PhysRevLett.62.324} {\bibfield
   {journal} {\bibinfo  {journal} {Phys. Rev. Lett.}\ }\textbf {\bibinfo
  {volume} {62}},\ \bibinfo {pages} {324} (\bibinfo {year} {1989})}\BibitemShut
  {NoStop}%
\bibitem [{\citenamefont {Jarrell}(1992)}]{DMFTJarrell1992}%
  \BibitemOpen
  \bibfield  {author} {\bibinfo {author} {\bibfnamefont {M.}~\bibnamefont
  {Jarrell}},\ }\href {https://doi.org/10.1103/PhysRevLett.69.168} {\bibfield
  {journal} {\bibinfo  {journal} {Phys. Rev. Lett.}\ }\textbf {\bibinfo
  {volume} {69}},\ \bibinfo {pages} {168} (\bibinfo {year} {1992})}\BibitemShut
  {NoStop}%
\bibitem [{\citenamefont {Georges}\ and\ \citenamefont
  {Kotliar}(1992)}]{Georges1992}%
  \BibitemOpen
  \bibfield  {author} {\bibinfo {author} {\bibfnamefont {A.}~\bibnamefont
  {Georges}}\ and\ \bibinfo {author} {\bibfnamefont {G.}~\bibnamefont
  {Kotliar}},\ }\href {https://doi.org/10.1103/PhysRevB.45.6479} {\bibfield
  {journal} {\bibinfo  {journal} {Phys. Rev. B}\ }\textbf {\bibinfo {volume}
  {45}},\ \bibinfo {pages} {6479} (\bibinfo {year} {1992})}\BibitemShut
  {NoStop}%
\bibitem [{\citenamefont {Georges}\ \emph
  {et~al.}(1996{\natexlab{a}})\citenamefont {Georges}, \citenamefont {Kotliar},
  \citenamefont {Krauth},\ and\ \citenamefont {Rozenberg}}]{DMFTRMP1996}%
  \BibitemOpen
  \bibfield  {author} {\bibinfo {author} {\bibfnamefont {A.}~\bibnamefont
  {Georges}}, \bibinfo {author} {\bibfnamefont {G.}~\bibnamefont {Kotliar}},
  \bibinfo {author} {\bibfnamefont {W.}~\bibnamefont {Krauth}},\ and\ \bibinfo
  {author} {\bibfnamefont {M.~J.}\ \bibnamefont {Rozenberg}},\ }\href
  {https://doi.org/10.1103/RevModPhys.68.13} {\bibfield  {journal} {\bibinfo
  {journal} {Rev. Mod. Phys.}\ }\textbf {\bibinfo {volume} {68}},\ \bibinfo
  {pages} {13} (\bibinfo {year} {1996}{\natexlab{a}})}\BibitemShut {NoStop}%
\bibitem [{\citenamefont {Kotliar}\ \emph
  {et~al.}(2006{\natexlab{a}})\citenamefont {Kotliar}, \citenamefont
  {Savrasov}, \citenamefont {Haule}, \citenamefont {Oudovenko}, \citenamefont
  {Parcollet},\ and\ \citenamefont {Marianetti}}]{DMFTRMP2006}%
  \BibitemOpen
  \bibfield  {author} {\bibinfo {author} {\bibfnamefont {G.}~\bibnamefont
  {Kotliar}}, \bibinfo {author} {\bibfnamefont {S.~Y.}\ \bibnamefont
  {Savrasov}}, \bibinfo {author} {\bibfnamefont {K.}~\bibnamefont {Haule}},
  \bibinfo {author} {\bibfnamefont {V.~S.}\ \bibnamefont {Oudovenko}}, \bibinfo
  {author} {\bibfnamefont {O.}~\bibnamefont {Parcollet}},\ and\ \bibinfo
  {author} {\bibfnamefont {C.~A.}\ \bibnamefont {Marianetti}},\ }\href
  {https://doi.org/10.1103/RevModPhys.78.865} {\bibfield  {journal} {\bibinfo
  {journal} {Rev. Mod. Phys.}\ }\textbf {\bibinfo {volume} {78}},\ \bibinfo
  {pages} {865} (\bibinfo {year} {2006}{\natexlab{a}})}\BibitemShut {NoStop}%
\bibitem [{\citenamefont {Georges}\ \emph
  {et~al.}(1996{\natexlab{b}})\citenamefont {Georges}, \citenamefont {Kotliar},
  \citenamefont {Krauth},\ and\ \citenamefont {Rozenberg}}]{RMPGeorges1996}%
  \BibitemOpen
  \bibfield  {author} {\bibinfo {author} {\bibfnamefont {A.}~\bibnamefont
  {Georges}}, \bibinfo {author} {\bibfnamefont {G.}~\bibnamefont {Kotliar}},
  \bibinfo {author} {\bibfnamefont {W.}~\bibnamefont {Krauth}},\ and\ \bibinfo
  {author} {\bibfnamefont {M.~J.}\ \bibnamefont {Rozenberg}},\ }\href
  {https://doi.org/10.1103/RevModPhys.68.13} {\bibfield  {journal} {\bibinfo
  {journal} {Rev. Mod. Phys.}\ }\textbf {\bibinfo {volume} {68}},\ \bibinfo
  {pages} {13} (\bibinfo {year} {1996}{\natexlab{b}})}\BibitemShut {NoStop}%
\bibitem [{\citenamefont {Kotliar}\ \emph
  {et~al.}(2006{\natexlab{b}})\citenamefont {Kotliar}, \citenamefont
  {Savrasov}, \citenamefont {Haule}, \citenamefont {Oudovenko}, \citenamefont
  {Parcollet},\ and\ \citenamefont {Marianetti}}]{RMPKotliar2006}%
  \BibitemOpen
  \bibfield  {author} {\bibinfo {author} {\bibfnamefont {G.}~\bibnamefont
  {Kotliar}}, \bibinfo {author} {\bibfnamefont {S.~Y.}\ \bibnamefont
  {Savrasov}}, \bibinfo {author} {\bibfnamefont {K.}~\bibnamefont {Haule}},
  \bibinfo {author} {\bibfnamefont {V.~S.}\ \bibnamefont {Oudovenko}}, \bibinfo
  {author} {\bibfnamefont {O.}~\bibnamefont {Parcollet}},\ and\ \bibinfo
  {author} {\bibfnamefont {C.~A.}\ \bibnamefont {Marianetti}},\ }\href
  {https://doi.org/10.1103/RevModPhys.78.865} {\bibfield  {journal} {\bibinfo
  {journal} {Rev. Mod. Phys.}\ }\textbf {\bibinfo {volume} {78}},\ \bibinfo
  {pages} {865} (\bibinfo {year} {2006}{\natexlab{b}})}\BibitemShut {NoStop}%
\bibitem [{\citenamefont {Müller-Hartmann}(1989)}]{Hartmann1989}%
  \BibitemOpen
  \bibfield  {author} {\bibinfo {author} {\bibfnamefont {E.}~\bibnamefont
  {Müller-Hartmann}},\ }\href {https://doi.org/10.1007/BF01312686} {\bibfield
  {journal} {\bibinfo  {journal} {Z. Physik B - Condensed Matter}\ }\textbf
  {\bibinfo {volume} {76}},\ \bibinfo {pages} {211} (\bibinfo {year}
  {1989})}\BibitemShut {NoStop}%
\bibitem [{\citenamefont {Gonis}(1992)}]{gonis1992green}%
  \BibitemOpen
  \bibfield  {author} {\bibinfo {author} {\bibfnamefont {A.}~\bibnamefont
  {Gonis}},\ }\href {https://books.google.com/books?id=SqrvAAAAMAAJ} {\emph
  {\bibinfo {title} {Green Functions for Ordered and Disordered Systems}}},\
  Studies in mathematical physics\ (\bibinfo  {publisher} {North-Holland},\
  \bibinfo {year} {1992})\BibitemShut {NoStop}%
\bibitem [{\citenamefont {Elliott}\ \emph {et~al.}(1974)\citenamefont
  {Elliott}, \citenamefont {Krumhansl},\ and\ \citenamefont {Leath}}]{CPA1974}%
  \BibitemOpen
  \bibfield  {author} {\bibinfo {author} {\bibfnamefont {R.~J.}\ \bibnamefont
  {Elliott}}, \bibinfo {author} {\bibfnamefont {J.~A.}\ \bibnamefont
  {Krumhansl}},\ and\ \bibinfo {author} {\bibfnamefont {P.~L.}\ \bibnamefont
  {Leath}},\ }\href {https://doi.org/10.1103/RevModPhys.46.465} {\bibfield
  {journal} {\bibinfo  {journal} {Rev. Mod. Phys.}\ }\textbf {\bibinfo {volume}
  {46}},\ \bibinfo {pages} {465} (\bibinfo {year} {1974})}\BibitemShut
  {NoStop}%
\bibitem [{\citenamefont {Soven}(1967)}]{CPASoven1967}%
  \BibitemOpen
  \bibfield  {author} {\bibinfo {author} {\bibfnamefont {P.}~\bibnamefont
  {Soven}},\ }\href {https://doi.org/10.1103/PhysRev.156.809} {\bibfield
  {journal} {\bibinfo  {journal} {Phys. Rev.}\ }\textbf {\bibinfo {volume}
  {156}},\ \bibinfo {pages} {809} (\bibinfo {year} {1967})}\BibitemShut
  {NoStop}%
\bibitem [{\citenamefont {Taylor}(1967)}]{CPATaylor1967}%
  \BibitemOpen
  \bibfield  {author} {\bibinfo {author} {\bibfnamefont {D.~W.}\ \bibnamefont
  {Taylor}},\ }\href {https://doi.org/10.1103/PhysRev.156.1017} {\bibfield
  {journal} {\bibinfo  {journal} {Phys. Rev.}\ }\textbf {\bibinfo {volume}
  {156}},\ \bibinfo {pages} {1017} (\bibinfo {year} {1967})}\BibitemShut
  {NoStop}%
\bibitem [{\citenamefont {Velick\'y}\ \emph {et~al.}(1968)\citenamefont
  {Velick\'y}, \citenamefont {Kirkpatrick},\ and\ \citenamefont
  {Ehrenreich}}]{CPA1968}%
  \BibitemOpen
  \bibfield  {author} {\bibinfo {author} {\bibfnamefont {B.}~\bibnamefont
  {Velick\'y}}, \bibinfo {author} {\bibfnamefont {S.}~\bibnamefont
  {Kirkpatrick}},\ and\ \bibinfo {author} {\bibfnamefont {H.}~\bibnamefont
  {Ehrenreich}},\ }\href {https://doi.org/10.1103/PhysRev.175.747} {\bibfield
  {journal} {\bibinfo  {journal} {Phys. Rev.}\ }\textbf {\bibinfo {volume}
  {175}},\ \bibinfo {pages} {747} (\bibinfo {year} {1968})}\BibitemShut
  {NoStop}%
\bibitem [{\citenamefont {Dohner}\ \emph {et~al.}(2022)\citenamefont {Dohner},
  \citenamefont {Terletska}, \citenamefont {Tam}, \citenamefont {Moreno},\ and\
  \citenamefont {Fotso}}]{CPADMFTnonequilibrium2022}%
  \BibitemOpen
  \bibfield  {author} {\bibinfo {author} {\bibfnamefont {E.}~\bibnamefont
  {Dohner}}, \bibinfo {author} {\bibfnamefont {H.}~\bibnamefont {Terletska}},
  \bibinfo {author} {\bibfnamefont {K.-M.}\ \bibnamefont {Tam}}, \bibinfo
  {author} {\bibfnamefont {J.}~\bibnamefont {Moreno}},\ and\ \bibinfo {author}
  {\bibfnamefont {H.~F.}\ \bibnamefont {Fotso}},\ }\href
  {https://doi.org/10.1103/PhysRevB.106.195156} {\bibfield  {journal} {\bibinfo
   {journal} {Phys. Rev. B}\ }\textbf {\bibinfo {volume} {106}},\ \bibinfo
  {pages} {195156} (\bibinfo {year} {2022})}\BibitemShut {NoStop}%
\bibitem [{\citenamefont {Yan}\ and\ \citenamefont
  {Werner}(2023)}]{CPADMFTnonequilibrium2023}%
  \BibitemOpen
  \bibfield  {author} {\bibinfo {author} {\bibfnamefont {J.}~\bibnamefont
  {Yan}}\ and\ \bibinfo {author} {\bibfnamefont {P.}~\bibnamefont {Werner}},\
  }\href {https://doi.org/10.1103/PhysRevB.108.125143} {\bibfield  {journal}
  {\bibinfo  {journal} {Phys. Rev. B}\ }\textbf {\bibinfo {volume} {108}},\
  \bibinfo {pages} {125143} (\bibinfo {year} {2023})}\BibitemShut {NoStop}%
\bibitem [{\citenamefont {Poteryaev}\ \emph {et~al.}(2016)\citenamefont
  {Poteryaev}, \citenamefont {Skorikov}, \citenamefont {Anisimov},\ and\
  \citenamefont {Korotin}}]{CPADMFTDFT2016}%
  \BibitemOpen
  \bibfield  {author} {\bibinfo {author} {\bibfnamefont {A.~I.}\ \bibnamefont
  {Poteryaev}}, \bibinfo {author} {\bibfnamefont {N.~A.}\ \bibnamefont
  {Skorikov}}, \bibinfo {author} {\bibfnamefont {V.~I.}\ \bibnamefont
  {Anisimov}},\ and\ \bibinfo {author} {\bibfnamefont {M.~A.}\ \bibnamefont
  {Korotin}},\ }\href {https://doi.org/10.1103/PhysRevB.93.205135} {\bibfield
  {journal} {\bibinfo  {journal} {Phys. Rev. B}\ }\textbf {\bibinfo {volume}
  {93}},\ \bibinfo {pages} {205135} (\bibinfo {year} {2016})}\BibitemShut
  {NoStop}%
\bibitem [{\citenamefont {Milovanovi\ifmmode~\acute{c}\else \'{c}\fi{}}\ \emph
  {et~al.}(1989)\citenamefont {Milovanovi\ifmmode~\acute{c}\else \'{c}\fi{}},
  \citenamefont {Sachdev},\ and\ \citenamefont {Bhatt}}]{offdiagonal1989}%
  \BibitemOpen
  \bibfield  {author} {\bibinfo {author} {\bibfnamefont {M.}~\bibnamefont
  {Milovanovi\ifmmode~\acute{c}\else \'{c}\fi{}}}, \bibinfo {author}
  {\bibfnamefont {S.}~\bibnamefont {Sachdev}},\ and\ \bibinfo {author}
  {\bibfnamefont {R.~N.}\ \bibnamefont {Bhatt}},\ }\href
  {https://doi.org/10.1103/PhysRevLett.63.82} {\bibfield  {journal} {\bibinfo
  {journal} {Phys. Rev. Lett.}\ }\textbf {\bibinfo {volume} {63}},\ \bibinfo
  {pages} {82} (\bibinfo {year} {1989})}\BibitemShut {NoStop}%
\bibitem [{\citenamefont {Ulmke}\ and\ \citenamefont
  {Scalettar}(1997)}]{offdiagonal1997}%
  \BibitemOpen
  \bibfield  {author} {\bibinfo {author} {\bibfnamefont {M.}~\bibnamefont
  {Ulmke}}\ and\ \bibinfo {author} {\bibfnamefont {R.~T.}\ \bibnamefont
  {Scalettar}},\ }\href {https://doi.org/10.1103/PhysRevB.55.4149} {\bibfield
  {journal} {\bibinfo  {journal} {Phys. Rev. B}\ }\textbf {\bibinfo {volume}
  {55}},\ \bibinfo {pages} {4149} (\bibinfo {year} {1997})}\BibitemShut
  {NoStop}%
\bibitem [{\citenamefont {Denteneer}\ \emph {et~al.}(2001)\citenamefont
  {Denteneer}, \citenamefont {Scalettar},\ and\ \citenamefont
  {Trivedi}}]{offdiagonal2001}%
  \BibitemOpen
  \bibfield  {author} {\bibinfo {author} {\bibfnamefont {P.~J.~H.}\
  \bibnamefont {Denteneer}}, \bibinfo {author} {\bibfnamefont {R.~T.}\
  \bibnamefont {Scalettar}},\ and\ \bibinfo {author} {\bibfnamefont
  {N.}~\bibnamefont {Trivedi}},\ }\href
  {https://doi.org/10.1103/PhysRevLett.87.146401} {\bibfield  {journal}
  {\bibinfo  {journal} {Phys. Rev. Lett.}\ }\textbf {\bibinfo {volume} {87}},\
  \bibinfo {pages} {146401} (\bibinfo {year} {2001})}\BibitemShut {NoStop}%
\bibitem [{\citenamefont {Weh}\ \emph {et~al.}(2021)\citenamefont {Weh},
  \citenamefont {Zhang}, \citenamefont {\"Ostlin}, \citenamefont {Terletska},
  \citenamefont {Bauernfeind}, \citenamefont {Tam}, \citenamefont {Evertz},
  \citenamefont {Byczuk}, \citenamefont {Vollhardt},\ and\ \citenamefont
  {Chioncel}}]{BEBDMFT2021}%
  \BibitemOpen
  \bibfield  {author} {\bibinfo {author} {\bibfnamefont {A.}~\bibnamefont
  {Weh}}, \bibinfo {author} {\bibfnamefont {Y.}~\bibnamefont {Zhang}}, \bibinfo
  {author} {\bibfnamefont {A.}~\bibnamefont {\"Ostlin}}, \bibinfo {author}
  {\bibfnamefont {H.}~\bibnamefont {Terletska}}, \bibinfo {author}
  {\bibfnamefont {D.}~\bibnamefont {Bauernfeind}}, \bibinfo {author}
  {\bibfnamefont {K.-M.}\ \bibnamefont {Tam}}, \bibinfo {author} {\bibfnamefont
  {H.~G.}\ \bibnamefont {Evertz}}, \bibinfo {author} {\bibfnamefont
  {K.}~\bibnamefont {Byczuk}}, \bibinfo {author} {\bibfnamefont
  {D.}~\bibnamefont {Vollhardt}},\ and\ \bibinfo {author} {\bibfnamefont
  {L.}~\bibnamefont {Chioncel}},\ }\href
  {https://doi.org/10.1103/PhysRevB.104.045127} {\bibfield  {journal} {\bibinfo
   {journal} {Phys. Rev. B}\ }\textbf {\bibinfo {volume} {104}},\ \bibinfo
  {pages} {045127} (\bibinfo {year} {2021})}\BibitemShut {NoStop}%
\bibitem [{\citenamefont {Dobrosavljevi\ifmmode~\acute{c}\else \'{c}\fi{}}\
  and\ \citenamefont {Kotliar}(1993)}]{randomhopping1993}%
  \BibitemOpen
  \bibfield  {author} {\bibinfo {author} {\bibfnamefont {V.}~\bibnamefont
  {Dobrosavljevi\ifmmode~\acute{c}\else \'{c}\fi{}}}\ and\ \bibinfo {author}
  {\bibfnamefont {G.}~\bibnamefont {Kotliar}},\ }\href
  {https://doi.org/10.1103/PhysRevLett.71.3218} {\bibfield  {journal} {\bibinfo
   {journal} {Phys. Rev. Lett.}\ }\textbf {\bibinfo {volume} {71}},\ \bibinfo
  {pages} {3218} (\bibinfo {year} {1993})}\BibitemShut {NoStop}%
\bibitem [{\citenamefont {Dobrosavljevi\ifmmode~\acute{c}\else \'{c}\fi{}}\
  and\ \citenamefont {Kotliar}(1994)}]{randomhopping1994}%
  \BibitemOpen
  \bibfield  {author} {\bibinfo {author} {\bibfnamefont {V.}~\bibnamefont
  {Dobrosavljevi\ifmmode~\acute{c}\else \'{c}\fi{}}}\ and\ \bibinfo {author}
  {\bibfnamefont {G.}~\bibnamefont {Kotliar}},\ }\href
  {https://doi.org/10.1103/PhysRevB.50.1430} {\bibfield  {journal} {\bibinfo
  {journal} {Phys. Rev. B}\ }\textbf {\bibinfo {volume} {50}},\ \bibinfo
  {pages} {1430} (\bibinfo {year} {1994})}\BibitemShut {NoStop}%
\bibitem [{\citenamefont {Cheng}\ \emph
  {et~al.}(2019{\natexlab{a}})\citenamefont {Cheng}, \citenamefont {Sang},
  \citenamefont {Zhai},\ and\ \citenamefont {Ke}}]{ACPAPhononband2019}%
  \BibitemOpen
  \bibfield  {author} {\bibinfo {author} {\bibfnamefont {Z.}~\bibnamefont
  {Cheng}}, \bibinfo {author} {\bibfnamefont {M.}~\bibnamefont {Sang}},
  \bibinfo {author} {\bibfnamefont {J.}~\bibnamefont {Zhai}},\ and\ \bibinfo
  {author} {\bibfnamefont {Y.}~\bibnamefont {Ke}},\ }\href
  {https://doi.org/10.1103/PhysRevB.100.214206} {\bibfield  {journal} {\bibinfo
   {journal} {Phys. Rev. B}\ }\textbf {\bibinfo {volume} {100}},\ \bibinfo
  {pages} {214206} (\bibinfo {year} {2019}{\natexlab{a}})}\BibitemShut
  {NoStop}%
\bibitem [{\citenamefont {Zhai}\ \emph {et~al.}(2021)\citenamefont {Zhai},
  \citenamefont {Xue}, \citenamefont {Cheng},\ and\ \citenamefont
  {Ke}}]{ACPAPhonon2021}%
  \BibitemOpen
  \bibfield  {author} {\bibinfo {author} {\bibfnamefont {J.}~\bibnamefont
  {Zhai}}, \bibinfo {author} {\bibfnamefont {R.}~\bibnamefont {Xue}}, \bibinfo
  {author} {\bibfnamefont {Z.}~\bibnamefont {Cheng}},\ and\ \bibinfo {author}
  {\bibfnamefont {Y.}~\bibnamefont {Ke}},\ }\href
  {https://doi.org/10.1103/PhysRevB.104.024205} {\bibfield  {journal} {\bibinfo
   {journal} {Phys. Rev. B}\ }\textbf {\bibinfo {volume} {104}},\ \bibinfo
  {pages} {024205} (\bibinfo {year} {2021})}\BibitemShut {NoStop}%
\bibitem [{\citenamefont {Cheng}\ \emph
  {et~al.}(2019{\natexlab{b}})\citenamefont {Cheng}, \citenamefont {Zhai},
  \citenamefont {Zhang},\ and\ \citenamefont {Ke}}]{ACPAPhonon2019}%
  \BibitemOpen
  \bibfield  {author} {\bibinfo {author} {\bibfnamefont {Z.}~\bibnamefont
  {Cheng}}, \bibinfo {author} {\bibfnamefont {J.}~\bibnamefont {Zhai}},
  \bibinfo {author} {\bibfnamefont {Q.}~\bibnamefont {Zhang}},\ and\ \bibinfo
  {author} {\bibfnamefont {Y.}~\bibnamefont {Ke}},\ }\href
  {https://doi.org/10.1103/PhysRevB.99.134202} {\bibfield  {journal} {\bibinfo
  {journal} {Phys. Rev. B}\ }\textbf {\bibinfo {volume} {99}},\ \bibinfo
  {pages} {134202} (\bibinfo {year} {2019}{\natexlab{b}})}\BibitemShut
  {NoStop}%
\bibitem [{\citenamefont {Zhai}\ \emph {et~al.}(2019)\citenamefont {Zhai},
  \citenamefont {Zhang}, \citenamefont {Cheng}, \citenamefont {Ren},
  \citenamefont {Ke},\ and\ \citenamefont {Li}}]{ACPAPhonontransport2019}%
  \BibitemOpen
  \bibfield  {author} {\bibinfo {author} {\bibfnamefont {J.}~\bibnamefont
  {Zhai}}, \bibinfo {author} {\bibfnamefont {Q.}~\bibnamefont {Zhang}},
  \bibinfo {author} {\bibfnamefont {Z.}~\bibnamefont {Cheng}}, \bibinfo
  {author} {\bibfnamefont {J.}~\bibnamefont {Ren}}, \bibinfo {author}
  {\bibfnamefont {Y.}~\bibnamefont {Ke}},\ and\ \bibinfo {author}
  {\bibfnamefont {B.}~\bibnamefont {Li}},\ }\href
  {https://doi.org/10.1103/PhysRevB.99.195429} {\bibfield  {journal} {\bibinfo
  {journal} {Phys. Rev. B}\ }\textbf {\bibinfo {volume} {99}},\ \bibinfo
  {pages} {195429} (\bibinfo {year} {2019})}\BibitemShut {NoStop}%
\bibitem [{\citenamefont {Wei}\ \emph {et~al.}(2022)\citenamefont {Wei},
  \citenamefont {Zhai}, \citenamefont {Ning},\ and\ \citenamefont
  {Ke}}]{Weiqi2022}%
  \BibitemOpen
  \bibfield  {author} {\bibinfo {author} {\bibfnamefont {Q.}~\bibnamefont
  {Wei}}, \bibinfo {author} {\bibfnamefont {J.}~\bibnamefont {Zhai}}, \bibinfo
  {author} {\bibfnamefont {Z.}~\bibnamefont {Ning}},\ and\ \bibinfo {author}
  {\bibfnamefont {Y.}~\bibnamefont {Ke}},\ }\href
  {https://doi.org/10.1103/PhysRevB.106.214205} {\bibfield  {journal} {\bibinfo
   {journal} {Phys. Rev. B}\ }\textbf {\bibinfo {volume} {106}},\ \bibinfo
  {pages} {214205} (\bibinfo {year} {2022})}\BibitemShut {NoStop}%
\bibitem [{\citenamefont {Zhai}\ \emph {et~al.}(2024)\citenamefont {Zhai},
  \citenamefont {Cheng}, \citenamefont {Zhang},\ and\ \citenamefont
  {Ke}}]{ACPAclustertheory}%
  \BibitemOpen
  \bibfield  {author} {\bibinfo {author} {\bibfnamefont {J.}~\bibnamefont
  {Zhai}}, \bibinfo {author} {\bibfnamefont {Z.}~\bibnamefont {Cheng}},
  \bibinfo {author} {\bibfnamefont {Y.}~\bibnamefont {Zhang}},\ and\ \bibinfo
  {author} {\bibfnamefont {Y.}~\bibnamefont {Ke}},\ }\href
  {https://doi.org/10.1103/PhysRevB.109.094203} {\bibfield  {journal} {\bibinfo
   {journal} {Phys. Rev. B}\ }\textbf {\bibinfo {volume} {109}},\ \bibinfo
  {pages} {094203} (\bibinfo {year} {2024})}\BibitemShut {NoStop}%
\bibitem [{\citenamefont {Cui}\ \emph {et~al.}()\citenamefont {Cui},
  \citenamefont {Zhang}, \citenamefont {Wei}, \citenamefont {Zhang},\ and\
  \citenamefont {Ke}}]{cui2024}%
  \BibitemOpen
  \bibfield  {author} {\bibinfo {author} {\bibfnamefont {R.}~\bibnamefont
  {Cui}}, \bibinfo {author} {\bibfnamefont {Z.}~\bibnamefont {Zhang}}, \bibinfo
  {author} {\bibfnamefont {Q.}~\bibnamefont {Wei}}, \bibinfo {author}
  {\bibfnamefont {Y.}~\bibnamefont {Zhang}},\ and\ \bibinfo {author}
  {\bibfnamefont {Y.}~\bibnamefont {Ke}},\ }\href@noop {} {\bibfield  {journal}
  {\bibinfo  {journal} {Phys. Rev. B}\ }}\bibinfo {note} {Submitted for
  publication}\BibitemShut {NoStop}%
\bibitem [{\citenamefont {Werner}\ \emph {et~al.}(2006)\citenamefont {Werner},
  \citenamefont {Comanac}, \citenamefont {de' Medici}, \citenamefont {Troyer},\
  and\ \citenamefont {Millis}}]{CTHYB2006PRL}%
  \BibitemOpen
  \bibfield  {author} {\bibinfo {author} {\bibfnamefont {P.}~\bibnamefont
  {Werner}}, \bibinfo {author} {\bibfnamefont {A.}~\bibnamefont {Comanac}},
  \bibinfo {author} {\bibfnamefont {L.}~\bibnamefont {de' Medici}}, \bibinfo
  {author} {\bibfnamefont {M.}~\bibnamefont {Troyer}},\ and\ \bibinfo {author}
  {\bibfnamefont {A.~J.}\ \bibnamefont {Millis}},\ }\href
  {https://doi.org/10.1103/PhysRevLett.97.076405} {\bibfield  {journal}
  {\bibinfo  {journal} {Phys. Rev. Lett.}\ }\textbf {\bibinfo {volume} {97}},\
  \bibinfo {pages} {076405} (\bibinfo {year} {2006})}\BibitemShut {NoStop}%
\bibitem [{\citenamefont {Vollhardt}\ \emph {et~al.}(2012)\citenamefont
  {Vollhardt}, \citenamefont {Byczuk},\ and\ \citenamefont
  {Kollar}}]{Vollhardt2012}%
  \BibitemOpen
  \bibfield  {author} {\bibinfo {author} {\bibfnamefont {D.}~\bibnamefont
  {Vollhardt}}, \bibinfo {author} {\bibfnamefont {K.}~\bibnamefont {Byczuk}},\
  and\ \bibinfo {author} {\bibfnamefont {M.}~\bibnamefont {Kollar}},\ }\bibinfo
  {title} {Dynamical mean-field theory},\ in\ \href
  {https://doi.org/10.1007/978-3-642-21831-6_7} {\emph {\bibinfo {booktitle}
  {Strongly Correlated Systems: Theoretical Methods}}},\ \bibinfo {editor}
  {edited by\ \bibinfo {editor} {\bibfnamefont {A.}~\bibnamefont {Avella}}\
  and\ \bibinfo {editor} {\bibfnamefont {F.}~\bibnamefont {Mancini}}}\
  (\bibinfo  {publisher} {Springer Berlin Heidelberg},\ \bibinfo {address}
  {Berlin, Heidelberg},\ \bibinfo {year} {2012})\ pp.\ \bibinfo {pages}
  {203--236}\BibitemShut {NoStop}%
\bibitem [{\citenamefont {Gull}\ \emph {et~al.}(2011)\citenamefont {Gull},
  \citenamefont {Millis}, \citenamefont {Lichtenstein}, \citenamefont
  {Rubtsov}, \citenamefont {Troyer},\ and\ \citenamefont
  {Werner}}]{RMPCTQMC2011}%
  \BibitemOpen
  \bibfield  {author} {\bibinfo {author} {\bibfnamefont {E.}~\bibnamefont
  {Gull}}, \bibinfo {author} {\bibfnamefont {A.~J.}\ \bibnamefont {Millis}},
  \bibinfo {author} {\bibfnamefont {A.~I.}\ \bibnamefont {Lichtenstein}},
  \bibinfo {author} {\bibfnamefont {A.~N.}\ \bibnamefont {Rubtsov}}, \bibinfo
  {author} {\bibfnamefont {M.}~\bibnamefont {Troyer}},\ and\ \bibinfo {author}
  {\bibfnamefont {P.}~\bibnamefont {Werner}},\ }\href
  {https://doi.org/10.1103/RevModPhys.83.349} {\bibfield  {journal} {\bibinfo
  {journal} {Rev. Mod. Phys.}\ }\textbf {\bibinfo {volume} {83}},\ \bibinfo
  {pages} {349} (\bibinfo {year} {2011})}\BibitemShut {NoStop}%
\bibitem [{\citenamefont {Si}\ \emph {et~al.}(1994)\citenamefont {Si},
  \citenamefont {Rozenberg}, \citenamefont {Kotliar},\ and\ \citenamefont
  {Ruckenstein}}]{EDQimao1994}%
  \BibitemOpen
  \bibfield  {author} {\bibinfo {author} {\bibfnamefont {Q.}~\bibnamefont
  {Si}}, \bibinfo {author} {\bibfnamefont {M.~J.}\ \bibnamefont {Rozenberg}},
  \bibinfo {author} {\bibfnamefont {G.}~\bibnamefont {Kotliar}},\ and\ \bibinfo
  {author} {\bibfnamefont {A.~E.}\ \bibnamefont {Ruckenstein}},\ }\href
  {https://doi.org/10.1103/PhysRevLett.72.2761} {\bibfield  {journal} {\bibinfo
   {journal} {Phys. Rev. Lett.}\ }\textbf {\bibinfo {volume} {72}},\ \bibinfo
  {pages} {2761} (\bibinfo {year} {1994})}\BibitemShut {NoStop}%
\bibitem [{\citenamefont {Caffarel}\ and\ \citenamefont
  {Krauth}(1994)}]{EDCaffarel1994}%
  \BibitemOpen
  \bibfield  {author} {\bibinfo {author} {\bibfnamefont {M.}~\bibnamefont
  {Caffarel}}\ and\ \bibinfo {author} {\bibfnamefont {W.}~\bibnamefont
  {Krauth}},\ }\href {https://doi.org/10.1103/PhysRevLett.72.1545} {\bibfield
  {journal} {\bibinfo  {journal} {Phys. Rev. Lett.}\ }\textbf {\bibinfo
  {volume} {72}},\ \bibinfo {pages} {1545} (\bibinfo {year}
  {1994})}\BibitemShut {NoStop}%
\bibitem [{\citenamefont {Capone}\ \emph {et~al.}(2007)\citenamefont {Capone},
  \citenamefont {de' Medici},\ and\ \citenamefont {Georges}}]{EDCapone2007}%
  \BibitemOpen
  \bibfield  {author} {\bibinfo {author} {\bibfnamefont {M.}~\bibnamefont
  {Capone}}, \bibinfo {author} {\bibfnamefont {L.}~\bibnamefont {de' Medici}},\
  and\ \bibinfo {author} {\bibfnamefont {A.}~\bibnamefont {Georges}},\ }\href
  {https://doi.org/10.1103/PhysRevB.76.245116} {\bibfield  {journal} {\bibinfo
  {journal} {Phys. Rev. B}\ }\textbf {\bibinfo {volume} {76}},\ \bibinfo
  {pages} {245116} (\bibinfo {year} {2007})}\BibitemShut {NoStop}%
\bibitem [{\citenamefont {Wilson}(1975)}]{RMPNRG1975}%
  \BibitemOpen
  \bibfield  {author} {\bibinfo {author} {\bibfnamefont {K.~G.}\ \bibnamefont
  {Wilson}},\ }\href {https://doi.org/10.1103/RevModPhys.47.773} {\bibfield
  {journal} {\bibinfo  {journal} {Rev. Mod. Phys.}\ }\textbf {\bibinfo {volume}
  {47}},\ \bibinfo {pages} {773} (\bibinfo {year} {1975})}\BibitemShut
  {NoStop}%
\bibitem [{\citenamefont {Bulla}\ \emph {et~al.}(2008)\citenamefont {Bulla},
  \citenamefont {Costi},\ and\ \citenamefont {Pruschke}}]{RMPNRG2008}%
  \BibitemOpen
  \bibfield  {author} {\bibinfo {author} {\bibfnamefont {R.}~\bibnamefont
  {Bulla}}, \bibinfo {author} {\bibfnamefont {T.~A.}\ \bibnamefont {Costi}},\
  and\ \bibinfo {author} {\bibfnamefont {T.}~\bibnamefont {Pruschke}},\ }\href
  {https://doi.org/10.1103/RevModPhys.80.395} {\bibfield  {journal} {\bibinfo
  {journal} {Rev. Mod. Phys.}\ }\textbf {\bibinfo {volume} {80}},\ \bibinfo
  {pages} {395} (\bibinfo {year} {2008})}\BibitemShut {NoStop}%
\bibitem [{\citenamefont {Huang}\ \emph {et~al.}(2015)\citenamefont {Huang},
  \citenamefont {Wang}, \citenamefont {Meng}, \citenamefont {Du}, \citenamefont
  {Werner},\ and\ \citenamefont {Dai}}]{Huang_2015}%
  \BibitemOpen
  \bibfield  {author} {\bibinfo {author} {\bibfnamefont {L.}~\bibnamefont
  {Huang}}, \bibinfo {author} {\bibfnamefont {Y.}~\bibnamefont {Wang}},
  \bibinfo {author} {\bibfnamefont {Z.~Y.}\ \bibnamefont {Meng}}, \bibinfo
  {author} {\bibfnamefont {L.}~\bibnamefont {Du}}, \bibinfo {author}
  {\bibfnamefont {P.}~\bibnamefont {Werner}},\ and\ \bibinfo {author}
  {\bibfnamefont {X.}~\bibnamefont {Dai}},\ }\href
  {https://doi.org/10.1016/j.cpc.2015.04.020} {\bibfield  {journal} {\bibinfo
  {journal} {Computer Physics Communications}\ }\textbf {\bibinfo {volume}
  {195}},\ \bibinfo {pages} {140–160} (\bibinfo {year} {2015})}\BibitemShut
  {NoStop}%
\bibitem [{\citenamefont {Gubernatis}\ \emph {et~al.}(1991)\citenamefont
  {Gubernatis}, \citenamefont {Jarrell}, \citenamefont {Silver},\ and\
  \citenamefont {Sivia}}]{ME1991}%
  \BibitemOpen
  \bibfield  {author} {\bibinfo {author} {\bibfnamefont {J.~E.}\ \bibnamefont
  {Gubernatis}}, \bibinfo {author} {\bibfnamefont {M.}~\bibnamefont {Jarrell}},
  \bibinfo {author} {\bibfnamefont {R.~N.}\ \bibnamefont {Silver}},\ and\
  \bibinfo {author} {\bibfnamefont {D.~S.}\ \bibnamefont {Sivia}},\ }\href
  {https://doi.org/10.1103/PhysRevB.44.6011} {\bibfield  {journal} {\bibinfo
  {journal} {Phys. Rev. B}\ }\textbf {\bibinfo {volume} {44}},\ \bibinfo
  {pages} {6011} (\bibinfo {year} {1991})}\BibitemShut {NoStop}%
\bibitem [{\citenamefont {Jarrell}\ and\ \citenamefont
  {Gubernatis}(1996)}]{ME1996}%
  \BibitemOpen
  \bibfield  {author} {\bibinfo {author} {\bibfnamefont {M.}~\bibnamefont
  {Jarrell}}\ and\ \bibinfo {author} {\bibfnamefont {J.}~\bibnamefont
  {Gubernatis}},\ }\href
  {https://doi.org/https://doi.org/10.1016/0370-1573(95)00074-7} {\bibfield
  {journal} {\bibinfo  {journal} {Physics Reports}\ }\textbf {\bibinfo {volume}
  {269}},\ \bibinfo {pages} {133} (\bibinfo {year} {1996})}\BibitemShut
  {NoStop}%
\bibitem [{\citenamefont {Huang}(2023)}]{HUANG2023}%
  \BibitemOpen
  \bibfield  {author} {\bibinfo {author} {\bibfnamefont {L.}~\bibnamefont
  {Huang}},\ }\href {https://doi.org/https://doi.org/10.1016/j.cpc.2023.108863}
  {\bibfield  {journal} {\bibinfo  {journal} {Computer Physics Communications}\
  }\textbf {\bibinfo {volume} {292}},\ \bibinfo {pages} {108863} (\bibinfo
  {year} {2023})}\BibitemShut {NoStop}%
\bibitem [{\citenamefont {Hubbard}(1963)}]{HubbardI1963}%
  \BibitemOpen
  \bibfield  {author} {\bibinfo {author} {\bibfnamefont {J.}~\bibnamefont
  {Hubbard}},\ }\href {https://doi.org/10.1098/rspa.1963.0204} {\bibfield
  {journal} {\bibinfo  {journal} {Proc. R. Soc. Lond. A}\ }\textbf {\bibinfo
  {volume} {276}},\ \bibinfo {pages} {238–257} (\bibinfo {year}
  {1963})}\BibitemShut {NoStop}%
\end{thebibliography}%
\end{document}